\documentclass[12pt,preprint]{aastex}
\usepackage{epsfig}

\def\ie{{\it i.e.}}
\def\eg{{\it e.g.}}

\def\lesssim{\mathrel{\hbox{\rlap{\hbox{\lower4pt\hbox{$\sim$}}}\hbox{$<$}}}}
\def\gtrsim{\mathrel{\hbox{\rlap{\hbox{\lower4pt\hbox{$\sim$}}}\hbox{$>$}}}}
\def\alt{\mathrel{\hbox{\rlap{\hbox{\lower4pt\hbox{$\sim$}}}\hbox{$<$}}}}
\def\agt{\mathrel{\hbox{\rlap{\hbox{\lower4pt\hbox{$\sim$}}}\hbox{$>$}}}}

\newcommand {\be}       {\begin{equation}}
\newcommand {\ee}       {\end{equation}}

\newcommand{\DIOm}      {\Delta I_{\Omega}}

\newcommand{\mue}{{\mu_{\rm e}}}

\begin{document}

\tighten

\title{The Crustal Rigidity of a Neutron Star, and 
       Implications for PSR 1828-11 and other Precession Candidates}
\author{Curt Cutler}
\affil{Max-Planck-Institut fuer Gravitationsphysik\\
Golm bei Potsdam, Germany; cutler@aei-potsdam.mpg.de}

\author{Greg Ushomirsky}
\affil{California Institute of Technology,\\
Pasadena, California; gregus@tapir.caltech.edu}


\author{Bennett Link}
\affil{Montana State University, \\ 
         Bozeman, Montana; blink@dante.physics.montana.edu }

\begin{abstract}
Like the Earth, a neutron star (NS) can undergo torque-free precession
because some piece $\Delta I_d$ of its inertia tensor remains tied to
the crust's principal axes, as opposed to following the crust's
angular velocity vector.  The (body-frame) precession frequency
$\nu_p$ is $\nu_s \Delta I_d/I_C$, where $\nu_s$ is the NS's spin
frequency and $I_C$ is the moment of inertia associated with the
crustal nuclei, plus any component of the star tightly-coupled to the
crust over a timescale less than the spin period.  For a spinning NS
with a relaxed crust, $\Delta I_d = b \Delta I_\Omega$, where $\Delta
I_\Omega$ is the rotational oblateness of a fluid star rotating at
spin frequency $\Omega$, and $b$ is the NS's {\em rigidity parameter}.
A previous estimate of $b$ by Baym \& Pines (1971) gives $b \sim
10^{-5}$ for typical NS parameters. Here we calculate the rigidity
parameter $b$, and show that it is $\sim 40$ times smaller than the
Baym-Pines estimate.  We apply this result to PSR 1828-11, an isolated
pulsar whose correlated timing residuals and pulse shape variations
provide strong evidence for precession with a 511-day period.  We show
that this precession period is $\sim 250$ times shorter than one would
expect, assuming that: 1) the crust is relaxed (except for the
stresses induced by the precession itself), and, 2) the NS possesses
no other source of stress that would deform its figure (\eg, a strong
magnetic field). We conclude that the crust must be under significant
stress to explain the precession period of PSR B1828-11; such stress
arises naturally as the star spins down. Assuming that crustal
shear stresses do set the precession period, the star's reference angular
velocity (roughly, the spin at which the crust is most relaxed)
is $\approx 40$ Hz (i.e., $\approx \sqrt{250}$ times faster 
than today's spin), and the weighted average of the  
crust's present spindown strain 
is $\bar\sigma^{sd}_{ave} \simeq 5 \times 10^{-5}$.

We briefly describe the implications of our improved $b$ calculation
for other precession candidates.

\end{abstract}

\keywords{pulsars: individual(PSR1828-11)---stars: neutron---stars: rotation}


\section{Introduction}\label{sec:intro}

Stairs et al.~(2000) recently reported strong evidence for free
precession in the radio pulsar PSR 1828-11.  This pulsar has period $P_s
= 0.405\,$s, $\dot P_s = 6.00 \times 10^{-14}$, a characteristic age
of $\tau_c = $0.11 Myr and an inferred B-field of $5.0\times
10^{12}$G. Timing residuals for PSR 1828-11 show strong periodic
modulation, with periods 511 and 256 days.  A 1009-day periodicity is
also claimed, but with lower confidence.  

The 511- and 256-day timing residuals are matched by periodic
modulations of the pulse shape, which strongly supports the
interpretation that the modulations are due to NS precession, with a
wobble amplitude of $\sim 3^\circ$. In the model of Link \& Epstein
(2001), free precession sets the 511-d timescale, while coupling of
the precession to the external electromagnetic torque produces
variations in the NS's spindown rate that give the main contribution
to the timing residuals (due to the changing angle between the neutron
star's magnetic dipole axis and spin direction) and produce a harmonic
at 256 d.

While periodic or quasi-periodic timing residuals have been observed in several other
isolated pulsars (including B1642-03, the Crab, and Vela) and
tentatively interpreted as evidence for torque-free precession (see
Jones \& Anderrson 2001 for a review), PSR 1828-11 certainly
represents the most convincing case for free precession in a neutron
star (NS).
This pulsar has prompted us to re-examine theoretical estimates of a
NS's (body-frame) free-precession frequency, $\nu_p$.  If the crust is
essentially relaxed, the body-frame precession period is $P_p =
(\nu_s)^{-1} I_C/\Delta I_d$, where $\nu_s$ is the spin frequency, $I_C$ is
the moment of inertia of the crust (plus any portion of the NS
strongly coupled to the crust on a timescale $<\nu_s^{-1}$), and
$\Delta I_d$ is the residual oblateness the NS {\it would} have if it
were spun down to zero frequency without the crust breaking or
otherwise relaxing (Pines \& Shaham 1972, Munk \& MacDonald 1960).  
For a crust that is relaxed at
its current spin rate, we can write \be\label{eq:bdef} \Delta I_d =
b\Delta I_\Omega \ , \ee
\noindent where $\Delta I_\Omega$ is the piece of NS's
inertia tensor due to its spin (i.e., to centrifugal force),
approximately given by (Jones 2000):
\be
\Delta I_\Omega \approx 0.3 I_0 (\nu_s/{\rm kHz})^2 \ ,
\ee
\noindent where $I_0$ is the total moment of inertia for the
star when non-rotating. The coefficient $b$ in
Eq.~(\ref{eq:bdef}) is sometimes called the NS's ``rigidity
parameter'' (Jones \& Andersson 2000).

This paper is concerned with an accurate calculation of $b$. For the
case of a star with constant density and constant shear modulus, the
result is $b =\frac{57}{10}\mu V/|E_g|$, where $\mu$ is the shear
modulus, $V$ is the star's volume, and $E_g = -\frac{3}{5} G M^2/R$ is
star's gravitational binding energy (Love 1944).  As an estimate for a
neutron star with a liquid core, Baym \& Pines (1971) took: 
\begin{equation}
\label{eq:bBP} 
b \approx \frac{57}{10}
|E_g|^{-1} \int{\mu\, dV} \sim 10^{-5}\ ,
\end{equation}
for a $1.4 M_\odot$ NS. The Baym-Pines estimate for $b$,
Eq.~(\ref{eq:bBP}), has been routinely adopted in the NS-precession 
literature.

In this paper, we mainly do two things.  First, we calculate $b$ for
realistic NS structure -- a solid crust afloat on a liquid core. We
solve for the strain field that develops as the NS spins down, and
find that $b$ is smaller than found by Baym and Pines by a factor of
$\sim 40$. A partial explanation of why Eq.~(\ref{eq:bBP}) yields so
large an overestimate is given in the Appendices.

Second, we use our improved $b$ value to show that PSR 1828-11 is
precessing $\sim 250$ times faster than one would expect, assuming its
crust is nearly relaxed.  Under the hypothesis that crust rigidity is
the dominant source of the deformation bulge $\Delta I_d$ (and that
the observed modulations are indeed due to precession), we conclude
that the crust of PSR 1828-11 is {\it not} currently relaxed, and that
its reference spin $\nu_{s,ref}$ 
(roughly, the spin at which the crust is most relaxed) is
approximately $40\,$Hz; i.e., $\approx \sqrt{250}$ times higher than
the current spin.  (In this sense, the precession of PSR B1828-11 is
different from the Chandler wobble of the Earth, since the Earth {\it
is} almost relaxed). 

As we will explain in \S 5, the current strain tensor in the crust
$\sigma_{ab}$ can be written as the sum of ``spindown'' (sd) and
``reference'' (ref) pieces: $\sigma_{ab} = \sigma^{sd}_{ab} +
\sigma^{ref}_{ab}$.  The $\sigma^{ref}_{ab}$ piece is basically the
strain that would exist in the crust, even if it were spinning at
$\nu_{s,ref}$; $\sigma^{ref}_{ab}$ depends on the detailed history of
local crust quakes/relaxations, and is thus unknown. The spindown
strain $\sigma^{sd}_{ab}$ is the strain induced as the NS spins down
from $\nu_{ref}$ to its current value $\nu_s$.  For PSR 1828-11, we
show that its weighted-average is $\bar \sigma^{sd}_{\rm ave} \sim 5
\times 10^{-5}$.

Although 
$\sigma^{ref}_{ab}$ is unknown, it seems likely to us
that the full strain is larger, on 
average, than the spindown strain.
(This is because the full strain $\bar\sigma \equiv \sqrt{0.5 \sigma_{ab} 
\sigma^{ab}}$ contains contributions from all
harmonics, while the spindown strain is entirely in the $Y_{20}$ harmonic.)
{\emph If} this is true, 
then the NS's crustal breaking strain $\bar\sigma_{max}$
must be larger than the 
average spindown value: $\bar\sigma_{max} > \bar \sigma^{sd}_{\rm
ave} \sim 5 \times 10^{-5}$.
Although $\bar\sigma_{max}$ is poorly constrained, both empirically and 
theoretically, a value this large seems quite reasonable: the usual 
assumption has been that 
$\bar\sigma_{max}$ for NS crusts is somewhere in the range $10^{-5}-10^{-1}$.
(Unlike the crust's shear modulus, which can
be estimated from simple energetics, the breaking strain $\bar \sigma_{max}$
depends on the type, density, and propagation of dislocations in 
the lattice, which are highly uncertain.)

If significant pinning of superfluid vortex lines to the nuclear
lattice or to flux tubes in the core occurs, the precession frequency
is not set by the material properties of the crust (Shaham 1977). In
this situation, the pinned superfluid behaves effectively as a large
$\Delta I_d$, equal to the moment of inertia $I_p$ of the pinned
superfluid.  For example, if the entire crust superfluid pins to the
solid, $I_p$ is $\sim 0.01 I_0 \sim I_C$. The (body-frame) precession
frequency should then be of order the spin frequency (or at most $\sim
100$ times slower, if the crust-core coupling is much larger than
usually presumed, so $I_C \sim I_0$).  Instead, in PSR 1828-11 and the
other precession candidates reviewed in \S 6, the precession frequency
is $\sim 6-8 $ orders of magnitude {\it lower} than the spin
frequency.  Thus claims for long-period NS precession appeared at
first to conflict with one explanation for large pulsar glitches,
wherein superfluid vortices do pin to crustal nuclei, and large
glitches represent large-scale unpinning events (see, {\sl e.g.},
Anderson \& Itoh 1975). This conflict was recently resolved by Link \&
Cutler (2001), who showed that, in PSR 1828-11 and other precession
candidates, the precessional motion itself exerts Magnus forces on the
vortices that are $\agt 100$ times larger than the Magnus forces
responsible for giant glitches. (The fact that sufficiently large,
precession-induced Magnus forces would unpin superfluid vortices had
been noted previously by Shaham 1986).  That is, in all likelihood the
precessional motion itself immediately unpins the superfluid vortices,
and keeps them unpinned.  Accordingly,
in this paper, we set $I_p = 0$. (Of course, pinning can persist
in precessing stars if the wobble angle is sufficiently small.)

This paper is concerned with whether crustal
stresses alone can give a NS precession period of order a year (for spin rates
of a few Hz), but 
we mention that magnetic field stresses can also give
a precession period of this order {\em if the core is a Type II
supeconductor}, as shown recently by Wasserman (2002), 
in an extension of work by
Mestel and collaborators (Mestel \& Takhar 1972; Mestel et al. 1981;
Nittman \& Wood 1981; see all Spitzer 1958).

The plan of this paper is as follows. In \S 2 we review the dynamics
of NS precession, partly to establish notation.  In \S 3 we present
our formalism for calculating the rigidity parameter $b$ for a NS.  In
\S 4 we present the results of this calculation: the value of $b$ and
the corresponding crustal stresses.  Knowing $b$ allows us to estimate
the precession frequency of a NS with relaxed crust. In \S 5 
we extend this result to a NS with substantially strained crust.  We
apply these results to PSR 1828-11 in \S 6, where we show that the
observed precession frequency is $\sim 250$ times higher than would be
predicted for a relaxed crust. We calculate the spindown strain levels
that must exist today in the crust of PSR 1828-11, assuming that
crustal shear stresses are indeed responsible for the observed
precession frequency.

In \S 7 we briefly discuss the implications of our work 
for some other NS's that show evidence for precession (the Crab, Vela,
PSR B1642-03, PSR 2217+47, PSR B0959-54, and the possible remnant in SN 1987A).
Our conclusions are summarized
in \S 8.

The Appendices represent our attempts to check and understand our
results for $b$ -- in particular to understand why the Baym-Pines
estimate is so much larger than the value we obtain.  In Appendix A we
represent $\Delta I_d$ as an integral over crustal shear stresses.
The result for $b$ we obtain in this way is consistent with our result
in \S 4, and gives more insight into how how different stress
components and different layers in the crust contribute to the final
answer.  In Appendix B we treat a case where $b$ can be obtained
analytically-- that of a uniform-density, incompressible star with a
thin crust of constant shear modulus afloat on a liquid core.  For
this case, we show the actual $b$ is $\simeq 5$ times smaller than the
Baym-Pines estimate, Eq.~(\ref{eq:bBP}).  In the realistic NS case,
$b$ is decreased by additional factor $\sim 8$, because the
contributions to $\Delta I_d$ from the different stress components and
from different layers in the crust tend to cancel each other much more
nearly than in the case of a thin, {\it uniform} crust. 

\section{Model and Results}\label{sec:results}
Here we derive the basic equations describing the precession of
a NS. In this section, and in \S\S 3-6, we will assume that
the only sources of non-sphericity in the NS are centrifugal
force and crustal shear stresses.

We idealize the precessing NS as having two components: C (for crust)
and F (for fluid), having angular velocities, $\Omega_C^a$ and
$\Omega_F^a$, respectively.  $\Omega_C^a$ and $\Omega_F^a$ have the
same magnitude, $\Omega$, but different directions. The $C$ piece
includes the crustal nuclei, but may also include some component of
the fluid that is coupled to the crust sufficiently strongly that it
effectively co-rotates with the crust over timescales shorter than the
rotation period $2\pi\Omega^{-1}$.  The F piece is that portion of the
fluid that is essentially decoupled from the crustal nuclei on this
timescale; conservation of vorticity then implies that the fluid
angular velocity $\Omega_F^a$ remains fixed, and so is aligned with
the NS's total angular momentum $J^a$: 
\be \Omega_F^a = \Omega \hat
J^a \ , \ee
\noindent where $\hat J^a$ is the unit vector along $J^a$. 
The angular velocity of the crustal nuclei is  $\Omega_C^a = \Omega 
\hat \Omega_C^a$, where $\hat \Omega_C^a$ is a unit vector.

We write the star's inertia tensor as
\begin{equation}
I_{ab}=I_{F,ab}+I_{C,ab}, 
\end{equation}
where $I_{F,ab}$ and $I_{C,ab}$ are the contributions from the fluid
interior and the solid crust, respectively. 
Following (and somewhat extending) Pines and Shaham (1972),
we approximate each contribution as 
the sum of a spherical piece, plus a 
centrifugal bulge that follows that component's angular velocity, 
plus a deformation bulge (sustained by crustal rigidity) that follows some
principal crustal axis $\hat n^a$: 

\begin{eqnarray}
 I_{F,ab} &=& I_{F,0}\delta_{ab} + \Delta I_{F,\Omega}(\hat J^a \hat J^b - \frac{1}{3} \delta_{ab})  \\ \nonumber
&+& \Delta I_{F,d} (\hat n_d^a \hat n_d^b   - \frac{1}{3} \delta_{ab} )
\label{IF} \\ 
 I_{C,ab} &=& I_{C,0}\delta_{ab} + \Delta I_{C,\Omega}(\hat\Omega_C^a \hat\Omega_C^b - \frac{1}{3} \delta_{ab})\\ \nonumber
&+& \Delta I_{F,d} (\hat n_d^a \hat n_d^b   - \frac{1}{3}  \delta_{ab}) \, . \label{IC}
\end{eqnarray}
\noindent where $\delta_{ab}$ is the unit tensor $[1,1,1]$.  The sum
$\Delta I_{d} \equiv \Delta I_{C,d} + \Delta I_{F,d}$ we call the NS's
{\em deformation bulge}; it is the non-sphericity the body would
retain if it were slowed down to zero angular velocity, without the
crust breaking or otherwise relaxing.  Although $\Delta I_{d}$ is
ultimately due to crustal shear stresses, the term $\Delta I_{F,d}$ is
non-zero (even in the absence of pinning), because the crust exerts
force on the fluid, both gravitationally and via pressure.

For a fully relaxed, rotating crust, we expect $\Delta I_d  \propto
\DIOm$, where $\DIOm \equiv \Delta I_{F,\Omega} + \Delta I_{C,\Omega}$ 
is the centrifugal piece of the entire star's inertia tensor, and the
coefficient of proportionality $b \equiv \Delta I_d/\DIOm$ is 
the rigidity parameter.
For a physical interpretation of $b$, let us take the Earth as an
example. The Earth's crust is essentially relaxed.  If we could slow
the Earth down to zero angular velocity without cracking its crust, it
would not settle directly into a spherical shape, but rather
would remain somewhat oblate. This is because the Earth's relaxed,
zero-strain shape is oblate, and after centrifugal forces are
removed, the stresses that build up in the crust will tend to push it back 
toward that relaxed shape. In fact, for the Earth, $b \approx 0.7$ 
(Munk and MacDonald 1960, p.40), so stopping the Earth from
spinning would reduce its oblateness by only $\sim 30\% $.  For a NS,
as we will see, $b \sim 2 \times 10^{-7}$, so halting a NS's rotation would 
decrease its oblateness by a factor of five million. The quantity
$b$ is so small for a NS because the gravitational energy
density far exceeds the crust's shear modulus.

The total angular momentum $J^a$ of fluid plus crust is
\be\label{eq:totang}
J_a = \Omega \bigl( I_{F,ab}\hat J^b  + I_{C,ab}\hat\Omega_C^b\bigr)
\ee
\noindent Note from Eqs.~(\ref{IF})-(\ref{eq:totang}) that
$\hat\Omega_C^a$, $\hat n^a$, and $\hat J^a$ are all coplanar.  Let $J^a$
lie along the z-axis, and let $\hat\Omega_C^a$ and $\hat n^a$ lie, at some
instant, in the $x-z$ plane, as illustrated in
Figure~\ref{fig:angles}.  The angle between the crust's symmetry
axis and angular velocity is $\theta$. The angle between the
angular velocity and the angular momentum is $\tilde\theta$.  These
angles are constants of the motion, and satisfy:
\begin{eqnarray}
\hat n^a = \cos\theta\hat z^a - \sin\theta\, \hat x^a \label{eq:n} \\
\hat\Omega_C^a = \cos\tilde\theta\hat z^a + \sin\tilde\theta\, \hat x^a  \, , \label{eq:Omc}
\end{eqnarray} 
From Eqs.~(\ref{eq:n})-(\ref{eq:Omc}) we have, up to third-order terms
in $\theta$ and $\tilde\theta$, 
\be
\label{eq:om_exp}
\hat\Omega_C^a = \Omega \left(1 + {\tilde\theta\over\theta} -
\frac{1}{2}\theta^2 -\frac{1}{2}\tilde\theta\theta\right)\hat J^a -
{\tilde\theta\over\theta} \Omega \,\hat n^a \, .  \ee From
Eq.~(\ref{eq:om_exp}), we see that the precession is a superposition
of two motions: $\hat n^a$ precesses around $\hat J^a$ with (inertial
frame) precession frequency $\dot\phi = \Omega(1 +
\tilde\theta/\theta)$, up to terms second order in $\theta$ and
$\tilde\theta$.  Likewise, the coefficient of $\hat n^a$ in
Eq.~(\ref{eq:om_exp}) gives the body-frame precession frequency:
$\Omega_p = \Omega (\tilde\theta/\theta)$. 
For $\Omega >0$ and
$\Delta I_d > 0$, an observer situated above the pole $\hat n^a$ and
fixed in the body frame sees the angular velocity $\hat \Omega^a$
circle around $\hat n^a$ in the counter-clockwise direction.
The ratio $\tilde\theta/\theta$ is
obtained directly by taking the $x$-component of Eq.~(\ref{eq:totang});
the result is 
\be 
\tilde\theta/\theta = {\Delta I_d\over I_{C,0} +
\frac{2}{3}\Delta I_{C,\Omega} - \frac{1}{3}\Delta I_{C,d} } \, , 
\ee
where $\Delta I_d \equiv \Delta I_{C,d} + \Delta I_{F,d}$. 
Since $\Delta I_{C,\Omega}$ and $\Delta I_{C,d}$ are tiny   
compared to $I_{C,0}$, we now ignore those terms
and drop the subscript ``0'' from $I_C$.
The (body 
frame) precession period $P_p$ is thus
\be
\label{eq:1}
P_p = P_s \ I_{C}/\Delta I_{d} \, .
\ee
\noindent 
Since the pulsar's magnetic dipole moment and emission region are
presumably fixed with respect to the crust, the angle between 
the dipole direction and
$\hat J^a$ varies on the timescale $P_p$; i.e., the body-frame precession 
period $P_p$ is the modulation period for the pulsed radio
signal.
Thus for PSR1828-11, the measured value of $\Delta I_d/I_C$ is  
\be
{\Delta I_d\over I_C} = (\nu_s P_p)^{-1} = 9.2 \times 10^{-9} 
\left ({511\mbox{ d}\over P_p}\right ). 
\ee
(We have 
included the factor $511 d/P_p$ explicitly in case the precession period 
is actually $\sim 1009 $ or $256$ days.)
\section{Elastic Deformation of the Slowing Crust: Formalism}
\label{sec:crust-deformation}
Our goal in \S 3-4 is to determine the rigidity parameter $b \equiv \Delta
I_d/\Delta I_\Omega$ for a NS with a relaxed crust. We first show
that $\Delta I_d$ can be re-expressed as the {\it difference} between 
(i) $\Delta I_\Omega$ for a purely fluid NS; and 
(ii) $\Delta I_\Omega$ for the same NS, but now having an elastic
crust whose relaxed state is spherical. We then describe how we 
solve for $\Delta I_\Omega$ for both cases (i) and (ii), using
perturbation theory.

Consider a 2-parameter family of NS's, all with the same mass and
composition, and all spinning about the $z$-axis. The two
parameters are the NS's actual angular velocity, $\Omega$, and 
the angular velocity $\Omega_r$ at which its crust is relaxed. 
For convenience, we change variables to 
$\Omega^2_r$ and $\Delta\Omega^2 \equiv
\Omega^2 -\Omega^2_r$.  
The star's inertia tensor $I_{ab}$ is a function of both
these variables: $I_{ab} = I_{ab}(\Omega^2_r, \Delta\Omega^2)$. By axial
symmetry, we can write 
\be\label{eq:full_expansion} 
I_{ab} = I_0(\Omega^2_r, \Delta\Omega^2)\delta_{ab} + I_2(\Omega^2_r,
\Delta\Omega^2) \bigl(\hat z^a \hat z^b - \frac{1}{3}\delta_{ab}\bigr) \, .
\ee
We define the function 
$\Delta I_d (\Omega)$ to be 
the ``residual'' $I_2$ of a NS that
is non-rotating, but whose crust is relaxed at angular velocity
$\Omega$.  Then, by definition, \be \Delta I_d(\Omega) \equiv
I_2(\Omega^2, -\Omega^2) \, , \ee
\noindent since $\Omega^2_r = \Omega^2$ means that 
the crust is relaxed at angular velocity $\Omega$, and
$\Delta\Omega^2 = -\Omega^2_r$ means that 
the star is non-rotating.

Note that $I_2(0,0) = 0$, since 
$\Omega^2_r = \Delta\Omega^2 = 0$
means both that the star's crust is relaxed and 
that the star is non-rotating, and hence the star is spherical.
Therefore we can write, to first order in 
$\Omega^2_r$ and $\Delta\Omega^2$,

\begin{eqnarray}
\Delta I_d(\Omega) &\equiv& I_2(\Omega^2, -\Omega^2)  \\
&=&  
\Omega^2 \frac{\partial I_2}{\partial (\Omega^2_r)}|_{0,0} -
\Omega^2 \frac{\partial I_2}{\partial (\Delta\Omega^2)}|_{0,0} \\
&=&  I_2(\Omega^2, 0) -  I_2(0,\Omega^2)  \ .
\end{eqnarray}

This suggests our strategy for computing $\Delta I_d (\Omega)$: 
we will compute both $I_2(\Omega^2, 0)$ and $I_2(0,\Omega^2)$
and take the difference.
Now, $I_2(\Omega^2,0)$ is the $I_2$ of a star
spinning at $\Omega$, whose crust is completely relaxed
at that spin. But if the crust is relaxed (i.e., if all shear stresses
vanish), the shape of the star is the same as for a completely fluid
star with the same mass and angular velocity.
On the other hand, $I_2(0,\Omega^2)$ is the $I_2$ of a star
whose angular velocity is $\Omega$, but whose
whose crust is relaxed when spherical (i.e., when it is nonspinning). 

Our computational strategy is to start with two nonspinning NS's, (i)
and (ii), identical except that (i) is treated as a fluid throughout,
while (ii) has an elastic crust whose relaxed shape is spherical.  Now
spin them both up to angular velocity $\Omega$, calculate $I_2$ for
each, and take the difference. The result is $\Delta I_d$ for a NS
whose crust is relaxed at spin $\Omega$.  (More precisely, we find
$\Delta I_d(\Omega)$, up to fractional corrections of order 
$\Omega^2 R^3/GM$.)
\subsection{The perturbation equations} 
The advantage to the strategy outlined above is that {\it both} 
problems (i) and (ii) can be 
treated with a linear analysis about a spherical
background for the stellar structure, 
with centrifugal force acting as a source term that drives the star
away from sphericity.
The effect of centrifugal force is described by the following
potential in spherical coordinates (with $\theta=0$ corresponding to
the rotation axis): 
\begin{equation}
\delta\phi^c=-\frac{1}{2}\Omega^2 r^2 \left(1-\cos^2\theta\right).
\end{equation}
This potential has both an $l=0$ and $l=2,m=0$ piece,
$\delta\phi^c=\delta\phi^c_0+\delta\phi^c_2$, where
\begin{equation}
\delta\phi^c_0=-\frac{1}{3}\Omega^2 r^2
\end{equation}
and
\begin{equation}
\delta \phi^c_2 (r,\theta,\phi)
= \hat\lambda Y_{20}(\theta,\phi) \Omega^2 r^2\, ,
\end{equation}
\noindent with \be \hat\lambda \equiv \frac{1}{3}\sqrt{\frac{4\pi}{5}}
\, .  \ee The $l=0$ piece causes spherically-symmetric expansion of
the star, while the $l=2$ piece generates the equatorial bulge.  Only
the $l=2$ piece contributes to $I_2$. However, displacements due to
both $l=0$ and $l=2$ pieces generate strains in the crust and, hence,
are important in determining the maximum $I_2$ spin-change sustainable
by the crust without cracking or otherwise relaxing.  Previous work on
spindown-induced crustal strain (e.g., Baym \& Pines 1971; Franco et
al. 2000) considered incompressible NS models, where this issue does
not arise, as the $l=0$ piece of the displacement vanishes for
incompressible models.  We now derive the equations governing the
response of the star to any $\delta \phi^c$, and then specialize to
the $l=2$ and $l=0$ pieces separately. Note that our treatment will be
purely Newtonian.

\subsubsection{NS response to centrifugal potential 
$\delta \phi^c$}

Inside the crust, the perturbed stress tensor (the sum of pressure and shear
stresses) can be written as
\begin{equation}
\delta\tau_{ab}=-\delta p\, g_{ab}+ 2\mu\,\sigma_{ab},
\end{equation}
where $\delta p$ is the pressure perturbation, $\mu(r)$ is the crust's shear
modulus, and $\sigma_{ab}$ is the crust's strain tensor, defined by 
\begin{equation}\label{eq:straindef}
\sigma_{ab}=\frac{1}{2}\left(\nabla_a\xi_b+\nabla_b\xi_a-
\frac{2}{3}g_{ab}\nabla^c\xi_c\right) \, .
\end{equation}
\noindent Here $\xi^a$ is the displacement of the crust away
from its relaxed state. The perturbation $\delta\tau_{ab}$ is
determined by the condition of hydro-elastic
balance, 
\begin{equation}\label{eq:hydro-balance-general}
\nabla^a\delta\tau_{ab}=\delta\rho\, g\, \hat{r}_b + \rho\nabla_b
\left(\delta\phi +\delta\phi^c_2\right),
\end{equation}
where $\hat r_b$ is the radial unit vector. 
The perturbation $\delta\phi$ of the gravitational potential obeys
\begin{equation}
\nabla^2\delta\phi=4\pi G\, \delta\rho.
\end{equation}
Finally, the density change arising from the displacement $\xi^a$
obeys the continuity equation,
\begin{equation}\label{eq:continuity}
\delta\rho=-\nabla^a(\rho\xi_a) \, .
\end{equation}

The following bundary conditions arise from
Eq.~(\ref{eq:hydro-balance-general}): 1) $\delta \tau_{r\perp}$
(where ``$\perp$'' refers to components orthogonal to the
radial vector $\hat r^a$) must go to zero at both the
crust-core boundary and the stellar surface (using the fact that shear
stresses vanish just outside the crust), and 2) $\delta \tau_{rr} = 0$
for perturbations that {\em break spherical symmetry} (since $\delta
p$ must vanish both in the liquid core--else fluid would flow--and
above the star).  These conditions can be succinctly expressed as
\begin{equation}\label{eq:bc0}
\delta \tau_{ab}\,\hat r^b = 0, 
\end{equation}
\noindent 
at both boundaries.  
(See Ushomirsky, Cutler \& Bildsten 2000 for a more
detailed discussion of these boundary conditions).

In the fluid core there are only two first-order equations to solve, for $\delta\phi$ and $d\,\delta\phi/dr$.
To derive them, first note that equilibrium between 
pressure, gravitational, and
centrifugal forces requires
\begin{equation}
\label{eq:equil}
\nabla_a \delta p = -\delta \rho \,\nabla_a \phi - 
\rho \nabla_a \left[\delta \phi + 
\delta \phi^c \right]\, .
\end{equation}
Projecting Eq.~(\ref{eq:equil}) 
along the horizontal and radial directions, we have
\begin{eqnarray}
\delta p(r) = -\rho\, \left[\delta \phi + 
\delta \phi^c \right]\, \label{eq:f1} \\ 
{d\,\delta p\over dr} = -\delta\rho\, g  -\rho\, {d\over dr}\left[\delta \phi + 
\delta \phi^c \right]\,  .  \label{eq:f2} 
\end{eqnarray}
Taking $d/dr$ of Eq.~(\ref{eq:f1}) and subtracting Eq.~(\ref{eq:f2}), we have
\begin{equation}\label{eq:delta-rho-liquid}
\delta \rho = g^{-1}\frac{d\rho}{dr}\, \left[\delta \phi + 
\delta \phi^c \right]\,
\end{equation}
\noindent so Poisson's equation becomes 
\begin{equation}\label{eq:delta-rho-liquid2}
\nabla^2 \delta \phi = 4\pi G (g^{-1}\frac{d\rho}{dr})\left[\delta \phi + 
\delta \phi^c \right]\, .
\end{equation}

\subsubsection{$l=2$ deformations}

We now seperate 
the above equations into spherical harmonics, thereby obtaining a 
system of first-order ODEs.  Our treatment follows McDermott et
al. (1988); see also Ushomirsky, Cutler, \& Bildsten (2000, hereafter
UCB).  Throughout this subsection, it is implicit that all scalar
perturbations have $Y_{20}$ angular dependence; we will 
generally suppress
explicit $(l,m)$ subscripts on perturbed quantities.

We begin by writing the displacement vector $\xi^a$ as the sum
of radial and tangential pieces: 
\begin{equation}\label{eq:xi-components}
\xi^a \equiv \xi^r(r) Y_{lm} \hat r^a + \xi^\perp(r) \beta^{-1}r
\nabla^aY_{lm}, 
\end{equation}
where $\hat r^a$ is the unit radial vector and $\beta \equiv \sqrt{l(l+1)}$.
Then inside the crust, there are 6 variables: 
\begin{eqnarray}
z_1\equiv\frac{\xi^r}{r},&\qquad& z_2 \equiv  \frac{\Delta\tau_{rr}}{p}
 =  \frac{\delta\tau_{rr}}{p} - z_1 \frac{d\ln p}{d\ln r} \\ \nonumber
z_3\equiv\frac{\xi^\perp}{\beta r},&\qquad&
z_4 \equiv \frac{\Delta \tau_{r\perp}}{\beta p} =
 \frac{\delta\tau_{r\perp}}{\beta p}  \\ \nonumber
z_5 \equiv \frac{\delta\phi}{c^2}\frac{R^2}{r^2}, 
&\qquad& z_6 \equiv  \frac{d z_5}{d\ln r} \, .
\end{eqnarray}
\noindent where $\delta$ and $\Delta$
refer to Eulerian and Lagrangian variations, respectively.  The
definitions of $z_1$-$z_4$ and $\delta\tau_{ab}$ are the same as in UCB. 
The above definition of $z_5$, with $r^2$ in the denominator, factors out
the dominant behavior of $\delta\phi/c^2$ near $r=0$. 

The perturbation equations are (see McDermott et al. 1988; UCB): 
\begin{mathletters}\label{eq:perts}
\begin{eqnarray}
\frac{dz_1}{d\ln r} &=& -\left(1+2\frac{\alpha_2}{\alpha_3}\right)z_1
        +\frac{1}{\alpha_3}z_2 
        + 6\frac{\alpha_2}{\alpha_3}z_3 , \\ 
\frac{dz_2}{d\ln r} &=& \left(\tilde U\tilde V-4\tilde V
                +12\Gamma\frac{\alpha_1}{\alpha_3}\right)z_1 
        +\left(\tilde V-4\frac{\alpha_1}{\alpha_3}\right)z_2 \\ \nonumber
        &+& 6\left(\tilde V-6\Gamma
         \frac{\alpha_1}{\alpha_3}\right)z_3
        +6 z_4 \\ \nonumber 
        &+&\frac{\tilde{V}c_3}{Z_\star}(2 z_5 + z_6)
        + 2 \frac{\tilde{V} c_3}{Z_\star}
        \hat{\lambda}\frac{R^2 \Omega^2}{c^2},  \\ 
\frac{dz_3}{d\ln r} &=& -z_1+\frac{1}{\alpha_1}z_4, \\
\frac{dz_4}{d\ln r} &=& 
        \left(\tilde V-6\Gamma\frac{\alpha_1}{\alpha_3}\right)z_1
        -\frac{\alpha_2}{\alpha_3}z_2  \\ \nonumber
        &+&\frac{2}{\alpha_3}\left\{11 \alpha_1\alpha_2
                +10\alpha_1^2\right\}z_3 \\ \nonumber
        &+&\left(\tilde V-3\right)z_4
        +\frac{\tilde{V}c_3}{Z_\star}z_5 +
        \frac{\tilde{V}c_3}{Z_\star}
        \hat\lambda\frac{R^2 \Omega^2}{c^2},  \\  
\frac{dz_5}{d\ln r} &=& z_6, \\
\frac{dz_6}{d\ln r} &=& - 5 z_6 + 
        \frac{\tilde{U} Z_\star}{c_3} \left[
\left(\frac{\tilde{V}}{\gamma}-4\frac{\alpha_1}{\alpha_3}\right)z_1
        \right. \\ \nonumber 
        &-& \left. \frac{1}{\alpha_3}z_2 
        +12 \frac{\alpha_1}{\alpha_3}z_3 \right], \\
\end{eqnarray}
\end{mathletters}
where 

\begin{eqnarray}\label{eq:defs}
\tilde U \equiv \frac{d\ln g}{d\ln r} + 2 ,  \ \ \tilde V
\equiv \frac{\rho \,g\,r}{p} = -\frac{d\ln p}{d\ln r}, \ \  
\alpha_1 \equiv \mu/p , \\ \nonumber
\Gamma=\left.\frac{\partial\ln p}{\partial\ln\rho}\right|_{\mue} \, ,\ \ \ \ 
\alpha_2 \equiv \Gamma - \frac{2}{3}\frac{\mu}{p} , 
\ \ \ \ \alpha_3 \equiv \Gamma +\frac{4}{3}\frac{\mu}{p} , \ \ \ \ 
\\ \nonumber
\gamma=\left.\frac{d\ln p}{d\ln\rho}\right|_{\star} \, , \ \ 
c_3=\left(\frac{r}{R}\right)^3 \left(\frac{M}{M_r}\right) \, , \ \  
Z_\star=\frac{GM}{Rc^2}.
\end{eqnarray}

In the fluid, we rewrite Eq.~(\ref{eq:delta-rho-liquid2}) in terms 
of our dimensionless
variables $z_5$ and $z_6$:
\begin{eqnarray}\label{eq:phi_core}
\frac{d z_5}{d\ln r} &=&  z_6 \, , \\
\frac{d z_6}{d\ln r} &=& -5 z_5 - \frac{\tilde{U}\tilde{V}}{\gamma}
\left(z_5 + \hat{\lambda}\frac{R^2\Omega^2}{c^2}
\right)\, .
\end{eqnarray}

The boundary conditions are as follows.   At the center of the star,
$d\delta\phi/dr$ must vanish, giving 
\begin{equation}
z_6=0,
\end{equation}
while $z_5$ takes on an (unknown) finite value. At both the crust-core
boundary ($r=r_c$) and at the top of the crust ($r=R$), we have, from
Eq.~({\ref{eq:bc0}),
\begin{eqnarray}\label{eq:bc2}
z_2 & = & \tilde{V} \left\{\frac{\rho_l}{\rho_s} z_1 
-\frac{c_3}{Z_\star} \left(z_5 +
\hat{\lambda}\frac{R^2\Omega^2}{c^2}\right)\right\}, \\
z_4 & = & 0. \nonumber
\end{eqnarray}
In addition, at the top of the crust ($r=R$), 
\begin{equation}\label{eq:outer}
5 z_5 + z_6 =0.
\end{equation}
Finally, at the crust-core boundary $\delta\phi$ and $d\delta\phi/dr$
must be continuous, requiring that $z_5$ and $z_6$ are as well. We
thus have 8 equations (two in the core and six in the crust) and 8
boundary conditions (one at the center, three at the surface, and four
at the crust-core boundary). For simplicity, we set the outer boundary
of the NS at the top of the crust. For convenience, we approximate
$\Gamma$ as $\gamma$ in Eqs.~(\ref{eq:perts}) (i.e., we do not
distinguish $d\ln p/d\ln\rho$ of the background model from the derivative at
constant composition). Solutions of these equations are described 
in \S~\ref{sec:results}.

The $I_2$ of the NS's inertia tensor is related to the
NS's quadrupole moment $Q_{20}$ by
\be\label{eq:I2Q2}
I_2 = \sqrt{\frac{4\pi}{5}} Q_{20} \, ,
\ee
where $Q_{20}$ is defined by 
\begin{equation}
Q_{20} \equiv \int \delta\rho(r) \ r^4 dr \, .
\end{equation}
\noindent 
$Q_{20}$ is related to $z_5$ at the NS surface by
\begin{equation}
\label{q20_tot}
Q_{20} = -\frac{5 c^2}{4\pi G}\, z_5|_{r=R} \, .
\end{equation}

As a check on the accuracy of our solutions, we also  
calculate $Q_{20}$ directly from the following integrals.
In the crust, $\delta\rho=-\nabla\cdot(\rho\vec{\xi})$, so 
\begin{eqnarray}
Q_{20}^{\rm crust} &=& 2\int_{r_c}^{R}\rho(z_1+3 z_3)r^4 dr -
\left[r^5 \rho z_1\right]^{r=R}_{r=r_c} \\ \nonumber
&=& \frac{MR^2}{4\pi}\left\{ 2\int_{r_c}^{R} \frac{\tilde{U}}{c_3}
\left(z_1+3z_3\right)\left(\frac{r}{R}\right)^5 d \ln r \right. \\ 
&-& \left.\left[\left(\frac{r}{R}\right)^5 \frac{\tilde{U}}{c_3} z_1
\right]^{r=R}_{r=r_c}\right\}\, . \label{eq:q2_crust}
\end{eqnarray}
In the liquid part of the star, we use Eq.~(\ref{eq:delta-rho-liquid})
to obtain
\begin{equation}\label{eq:q2_fluid}
Q_{20}^{\rm fluid}= - \frac{M R^2}{4\pi Z_\star}
\int_0^{r_c} \frac{\tilde{U}\tilde{V}}{\gamma}\left(
z_5 + \hat\lambda\frac{R^2\Omega^2}{c^2}\right)
\left(\frac{r}{R}\right)^5 d \ln r.
\end{equation}
In a star with both crust and core, we take $Q_{20}=Q_{20}^{\rm
fluid}+Q_{20}^{\rm crust}$. In practice, we refine our numerical solution 
until the sum of (\ref{eq:q2_crust}) and (\ref{eq:q2_fluid}) agrees
with (\ref{q20_tot}) to within required tolerance--about one part in
$10^{10}$.

To find $Q_{20}$ for a purely fluid star, we solve the same fluid
equations, but instead of matching at the bottom of the crust, we
impose the boundary condition Eq.~(\ref{eq:outer}) at $r=R$.  Neither
solution (i) nor (ii) is really accurate to one part in $10^{10}$,
because we approximate the star's outer boundary as being
at the top of the crust (ignoring the thin ocean on top), and because
we approximate $\Gamma$ by $\gamma$ in Eqs.~~(\ref{eq:perts}).
However since we use the same approximations in the solutions to
both problems (i) and (ii), the {\it difference} is due solely
to the presence of an elastic crust in model (ii). Therefore we claim
we do know difference between (i) and (ii) to three significant figures 
despite the fact that the two quantities agree to seven significant
figures. 
This claim is checked in Appendix A, where we obtain an
approximate analytical expression for $b$. 

\subsubsection{$l=0$ deformations}

The $l=0$ component of the centrifugal potential, $\delta\phi^c_0$,
generates a spherically symmetric radial displacement field
$\xi^a =\xi^r(r) \hat r^a$, as well as pressure, density, and
gravitational potential perturbations.  In this subsection, 
all scalar perturbed quantities are purely functions of $r$. 

In order to compute
$\xi^r$ for the $l=0$ case, we neglect the small shear modulus of the crust
(effectively taking the entire star to be fluid).   Of course, it is the small
shear modulus that determines $\Delta I_d$, and so in the previous
subsection our equations included it. However, our present purpose
is only to determine (roughly) the $l=0$ part of the crustal strain resulting
from NS spindown, and for this we can neglect $\mu$, since it has only 
a small influence on the radial displacement. This approximation is
justified by the smallness of the crust's shear modulus compared to
the pressure.
With this approximation, the perturbations in pressure, density,
and gravitational potential obey
\begin{eqnarray}
\frac{d\, \delta p}{dr}&=& -\delta\rho\, g -\rho \frac{d}{dr}
\left(\delta\phi+\delta\phi^c_0\right) \\
\frac{\delta p}{p}&=&\Gamma\frac{\delta\rho}{\rho}+
\left(\gamma-\Gamma\right)\xi^r\frac{d\ln\rho}{dr} \label{eq:dp0}\\
\nabla^2\delta\phi &=& 4\pi G\, \delta\rho \, .
\end{eqnarray}
Eq.~(\ref{eq:dp0}) comes from setting $\frac{\Delta p}{p} = \Gamma \frac{\Delta\rho}{\rho}$.

We define
dimensionless variables
\begin{eqnarray}
y_1&\equiv& \frac{\xi^r}{r}, \qquad 
y_2\equiv\frac{\delta p}{p} \\
\nonumber
y_3&\equiv& \frac{\delta\phi}{c^2}, \qquad 
y_4\equiv\frac{dy_3}{d\ln r}
\end{eqnarray}
and obtain 
\begin{mathletters}
\begin{eqnarray}
\frac{dy_1}{d\ln r} &=& \left(\frac{\tilde{V}}{\gamma}-3\right)y_1 -
\frac{1}{\gamma}y_2 \\
\frac{dy_2}{d \ln r} &=& \tilde{V}\left(1-\frac{1}{\gamma}\right)y_2 
-\frac{\tilde{V}c_3}{Z_\star}\left\{y_4 \left(\frac{R}{r}\right)^2
-\frac{2}{3}\frac{R^2\Omega^2}{c^2}\right\} \\
\frac{dy_3}{d\ln r} &\equiv& y_4 \\
\frac{dy_4}{d\ln r} &\equiv& \frac{\tilde{U}Z_\star}{c_3\gamma}
\left(\frac{r}{R}\right)^2y_2-y_4.
\end{eqnarray}
\end{mathletters}
Boundary conditions at $r=0$ are obtained from the requirement that
the solution be regular there.  This yields
\begin{equation}
\left(4-\frac{\Gamma}{\gamma}\right)y_1+\frac{1}{\Gamma}y_2 = 0 
\quad {\rm and}\quad y_4=0.
\end{equation}
At $r=R$, $\Delta p$ must vanish and the gravitational potential 
must match onto its solution in empty space 
($\delta\phi\propto 1/r$), so there
\begin{equation}
y_2-\tilde{V}y_1=0 \qquad {\rm and}\qquad y_3+y_4=0 \, , 
\end{equation}
while $y_3$ and $y_4$ are continuous at the crust-core interface.
\section{Elastic Deformation of the Slowing Crust: Results}
\label{sec:results} 
In this section, we describe the solutions to the elastic perturbation
equations derived in \S~\ref{sec:crust-deformation}, and 
present our results for the value of $b=\Delta I_d/I_{\Omega}$ and compare it
to the Baym-Pines estimate, Eq.~(\ref{eq:bBP}).  

In order to solve the perturbation equations, we need to specify the
background NS model.  We build NS models using several representative
equations of state (EOS), summarized in Table~\ref{table:eos-summary}.
These models vary somewhat in their treatment of the pressure-density
relation in the core and the crust.  In particular, model FPS uses the
equation of state described in Lorenz, Ravenhall \& Pethick (1993);
model WS uses EOS UV14+TNI of Wiringa et al. (1988)~\cite{Wiringa_88}
in the core, matched to model FPS in the crust.  Models AU and UU use
equations of state AV14+UVII and UV14+UVII, respectively, matched to
EOS of Negele \& Vautherin (1973) in the crust. In
practice, we use tabulations of pressure versus density provided with
the code RNS\footnote{N. Stergioulas,
http://www.gravity.phys.uwm.edu/rns/}.  Finally, model n1poly uses
pressure-density relation of an $n=1$ polytrope throughout the star.
In all cases, we assume zero temperature, take $M=1.4 M_\odot$, and
construct the NS model by solving the purely Newtonian equations of
hydrostatic balance, $dp/dr = - \rho g$, and continuity, $dM_r/dr=4\pi
r^2\rho$.  The resulting NS parameters (radius, $I_0$, $I_C$, and
$\Delta I_{\Omega}/\Omega^2$) are shown in Table~\ref{table:eos-summary}.

In this paper, we generally assume that $I_C$ equals the moment of inertia of the crustal nuclei
(i.e., excluding the dripped neutrons in the inner crust), which is  
roughly $1\%$ of $I_0$,  since
estimates of frictional coupling give coupling times between
the crust and core well in excess of the spin period (see, e.g., Alpar
\& Sauls 1988). If the stellar magnetic field penetrates the core,
magnetic stresses will also play a role in the coupling (Abney,
Epstein \& Olinto 1996; Mendell 1998), but the magnetic coupling timescale
(estimated as the Alfv\'en crossing time for waves in the core) 
is again much longer than the rotation period for most neutron stars.
However it is possible that some process {\it does} strongly couple
the core to crust (in which case $I_C$ could be a substantial fraction of
$I_0$), so we shall indicate how our main results scale with $I_C/I_0$.

In all models, we place the crust-core boundary at the fiducial
density of $\rho_b = 1.5\times10^{14}$~g~cm$^{-3}$ (Lorenz, Ravenhall
\& Pethick 1993). The crust-ocean boundary is
placed at $\rho=10^{9}$~g~cm$^{-3}$ for all models except for the
$n=1$ polytrope one, where it is placed at $10^{11}$~g~cm$^{-3}$ in
order to ensure numerical stability of our integration of the elastic
equations\footnote{In the $n=1$ polytrope model, the density falls of very
rapidly with radius in the outer layers--much faster than in models
with a more realistic EOS. This very rapid variation
over many orders of magnitude causes numerical problems. However, this
truncation does not affect the final results, as the outer crust
contains a negligible fraction of the mass.}.

We use the shear modulus $\mu$ as computed by Strohmayer et al.~(1991), via
Monte-Carlo simulations of both bcc crystals and quenched
solids. Their results can be conveniently rewritten in terms of
the pressure $p_e$ of degenerate relativistic electrons,
\begin{equation}\label{eq:shearmodulus}
\frac{\mu}{p_e}=\frac{6\times10^{-3}}{1+0.595(173/\Gamma_{\rm Coul})^2}
	\left(\frac{Z}{8}\right)^{2/3},
\end{equation}
where $\Gamma_{\rm Coul}={Z^2e^2}/{akT}$ is the ratio of the Coulomb
energy of the lattice to the thermal energy and $Z$ is the ionic
charge.  Throughout most of the crust (i.e., except near the top)
$\Gamma_{\rm Coul} > 10^{3}$, so we for simplicity we ignore the
slight dependence of the shear modulus on temperature, effectively
setting $1/\Gamma_{\rm Coul}$ equal to $0$. In order to compute the
composition ($A$ and $Z$ of nuclei, as well as the free neutron
fraction) we use the fits of Kaminker et al. (1999) to the formalism
of Oyamatsu (1993).  The resulting run of $\mu$ with density is shown
in Figure~\ref{fig:mu}.  For the $n=1$ polytrope model we use a
fiducial value of $\mu/p=10^{-2}$.

With the background model specified as above, we solve the
perturbation equations of \S~\ref{sec:crust-deformation} using
relaxation with adaptive mesh allocation (see, e.g., Press et
al. 1992~\cite{Press_92} for a description). For the case of $l=2$
perturbations in a purely fluid star, as well as for $l=0$
perturbations, the solution is straightforward. However, for $l=2$
perturbations in a star with a crust, there are internal boundary
conditions at the crust-core interface (Eq.~\ref{eq:bc2}), which makes
application of the standard relaxation method more complicated. In
this case, to deal with this internal boundary, we transform the
independent variable from $x=\ln r$ to $t$, such that for the fluid
part of the star $x=(x_c-x_0)t+x_0$, while in the crust
$x=x_1-(x_1-x_c)t$, where $x_0$, $x_1$, and $x_c$ are the values near
the center, at the crust-ocean boundary, and at the crust-core
boundary, respectively.  With this change of the independent variable,
{\it both} the fluid and crust equations can be solved simultaneously
on the interval $t\in[0,1]$: the internal boundary conditions are
eliminated.

In Figure~\ref{fig:z1} we show the radial and transverse displacements
induced when a relaxed, nonrotating star is spun up to $\Omega = c/R$.  
(This is the solution of our perturbation equations with 
$(R\Omega/c)^2$ set equal to $1$ in our source term Eq.~20.)
Of course, $\Omega = c/R$ is an unreasonably large spin value, 
but our solutions scale linearly
with the source term, \ie, quadratically in $\Omega$, so in all
applications we simply scale the results down to low $\Omega$.
Recall that, when solving the elastic perturbation equations, we
imagine taking a spherically symmetric star and spinning it up. In
this case, the $l=0$ part of the centrifugal force acts radially
outward, and hence the $l=0$ component of the radial displacement
(lower panel of Figure~\ref{fig:z1}) is positive throughout the crust.
On the other hand, the $l=2$ part of the centrigual force squeezes the
crust at the poles and pushes it out at the equator.  Hence, $z_1$
(top panel of Figure~\ref{fig:z1}), which can be thought of as the
radial displacement at the poles (times $R^{-1}\sqrt{4\pi/5}$), is negative. 
The transverse
displacement, $z_3$ (middle panel Figure~\ref{fig:z1}), changes sign
in order to satisfy mass continuity.

Adding the $l=0$ and $l=2$
perturbations, we can rewrite the strain tensor
(Eq.~\ref{eq:straindef}) as
\begin{equation}
\label{eq:sigab}
\sigma_{ab}=(\sigma_0+\sigma_{rr}Y_{20})(\hat{r}_a\hat{r}_b
-\frac{1}{2}e_{ab})
+\sigma_{r\perp}f_{ab} 
+\sigma_\Lambda(\Lambda_{ab}+\frac{1}{2}e_{ab}) \, .
\end{equation}
Here the tensors $e_{ab}$, $f_{ab}$, and $\Lambda_{ab}$ are defined by
\begin{mathletters}\label{eq:dtab-term-abbreviations}
\begin{eqnarray}
e_{ab} &\equiv& g_{ab} - \hat r_a\hat r_b, \\
f_{ab} &\equiv& \beta^{-1} r \bigl(\hat r_a \nabla_b Y_{lm} + 
\hat r_b \nabla_a Y_{lm})\bigr), \\
\Lambda_{ab} &\equiv& \frac{r^2}{l(l+1)}\nabla_a \nabla_b Y_{lm} +
f_{ab}\, ,
\end{eqnarray}
\end{mathletters}
\noindent the  $l=2$ strain components
$\sigma_{rr}$, $\sigma_{r\perp}$, and $\sigma_\Lambda$ are given by
\begin{mathletters}\label{eq:sigma-ab}
\begin{eqnarray}
\sigma_{rr}(r) &=& \frac{2}{3}\frac{dz_1}{d\ln r}+\frac{1}{3}\beta^2 z_3 \\
\sigma_{r\perp}(r) &=& \frac{\beta^2 z_4}{2\mu} \\
\sigma_{\Lambda}(r)&=& \beta^2 z_3 
\end{eqnarray}
\end{mathletters}
\noindent  and the $l=0$ piece is
\begin{equation}
\sigma_0=\frac{2}{3}\frac{dy_1}{d\ln r}
\end{equation}
In Figure~\ref{fig:sigma-all} we show the radial functions
$\sigma_{rr}$ (top panel), $\sigma_{r\perp}$ (second panel),
$\sigma_\Lambda$ (third panel), and $\sigma_0$ (bottom panel),
for equations of state AU, WS, and
n1poly. Again, these results are normalized to $(R\Omega/c)^2=1$. 

The crust breaks (or deforms plastically) when
\begin{equation}
\bar{\sigma} \equiv \biggl(\frac{1}{2}\sigma_{ab}\sigma^{ab}\biggr)^{1/2}
> \bar{\sigma}_{\rm max}, 
\end{equation}
where $\bar{\sigma}_{\rm max}$ is the {\em yield strain}. (This is the
von Mises criterion; see \S 6 of UCB for a discussion.)  From
Eq.~(\ref{eq:sigab}), we get
\begin{equation}
2\bar{\sigma}^2=\frac{3}{2}\left(\sigma_0+\sigma_{rr} Y_{20}\right)^2
+\sigma_{r\perp}^2 \frac{5}{64\pi}\left(\sin 2\theta\right)^2
+\sigma_\Lambda\frac{5}{32\pi}\sin^4\theta,
\end{equation}
where $Y_{20}=(5/16\pi)^{1/2}(3\cos^2\theta-1)$.  $\bar\sigma$ is
independent of $\phi$ by symmetry.  In Figure~\ref{fig:sigma-2d} we
plot contours of $\bar{\sigma}$ on a meridional plane (i.e., plane
that slices through the crust of the star at a constant longitude
$\phi$), and in Figure~\ref{fig:tau-2d} we show $\bar\tau=
2\mu\bar\sigma$, again on a meridional plane.  Clearly, the stresses
are largest at the base of the crust, where $\mu$ is largest, while
the strains are highest at the top of the crust (where they cost the
least, energetically).

In Table~\ref{table:eos-summary} we show the value of $b$ computed as
described in \S~\ref{sec:crust-deformation} using the solutions to the
perturbation problem and compare them to the estimates of Baym \&
Pines (1971) given by Eq.~(\ref{eq:bBP}). Eq.~(\ref{eq:bBP})
overestimates $b$ by a factor of $15-50$.  To emphasize this
difference, and also to explore its dependence on $\rho_b$ (the
assumed density at the crust-core boundary), we show in
Figure~\ref{fig:b-vs-stdguess} the ratio of our calculated $b$ to
the Baym-Pines estimate as a function of $\rho_b$.
In the two Appendices, we explore why the estimate Eq.~(\ref{eq:bBP})
is inaccurate. Briefly, when a uniform elastic sphere 
spins down, its strain energy is strongly concentrated toward
the center of the star; thus this case  (on which
Eq.~\ref{eq:bBP} is based) is unsuitable for estimating the
rigidity of a realistic NS, where all the strain energy is
near the surface, in the thin crust. Moreover, we show
that in a realistic NS crust, contributions to $\Delta I_d$ from
different stress components and different depths cancel each
other to a much higher degree than one might expect 
from consideration of a uniform, incompressible crust.

\section{$\Delta I_d$ for non-relaxed crusts}\label{gen}
So far in this paper we have calculated the residual oblateness
$\Delta I_d$ that a spinning NS with a {\it relaxed} crust retains if
it is gently torqued down to zero spin frequency. But a real 
NS that is spinning down (or spinning up, due to accretion) is probably 
not completely relaxed. Indeed, we shall see below that the
precession frequency of PSR1828-11 is inconsistent with the
assumption of a relaxed crust (assuming shear stresses are indeed 
responsible for this NS's $\Delta I_d$).
The crust, because it has finite rigidity, becomes strained 
as its spins changes. (As shown in Fig.~4 and \S 6, 
if no cracking or relaxation 
occurs, then the strain near the top of the crust grows to rather 
large values; e.g., for PSR 1828-11, the strain near the 
top would be $\sim 10^{-3}$ in this case.)
In some portions of the crust, 
the strain could be relaxed in localized 
episodes of crust cracking (starquakes) or plastic
flow, while in other regions the strain may simply build up.
Precisely because the NS crust may have a complex history
of strain build-up and release, 
its current strain pattern is impossible to predict
exactly.

Fortunately, however, many of our results can immediately be
generalized to this case of a non-relaxed crust, by scaling them to
the crust's ``reference spin'' $\nu_{s,ref}$, defined as follows.  We
imagine adjusting the NS's spin frequency $\nu_s$ while keepiing the
crust's preferred (zero-strain) shape fixed (i.e., without letting the
crust crack or otherwise relax). While in general there will be no
spin at which the crust is completely relaxed, if we plot the
resulting quadrupole moment $Q_{20}$ as a function of $\nu_s$, there
would be one spin value, $\nu_{s,ref}$, such that
$Q_{20}(\nu_{s,ref})$ was identical to the $Q_{20}$ of a perfect-fluid
NS with the same mass and spin, $\nu_{s,ref}$. (Here we are making the
natural assumption that the crust's preferred shape is roughly oblate,
as opposed to prolate.)  We define the crustal strain field at
$\nu_{s,ref}$ to be $\sigma^{ref}_{ab}$.  While $\sigma^{ref}_{ab}$ is
in general non-zero, these strains make no net contribution to the
NS's quadrupole moment. In a rough way, one can
think of $\nu_{s,ref}$ as the the spin at which the crust is {\em most
relaxed.} As an example, consider a NS that is born with a relaxed
crust at spin $\nu_{s,init}$ and never relaxes at all as it spins down
to $\nu_{s,final}$. Then $\nu_{s,ref} = \nu_{s,init}$ and
$\sigma^{ref}_{ab} = 0$.  If, on the other hand, the NS relaxes
somewhat (but not completely) as it spins down, then $\nu_{s,ref}$
will lie somewhere between $\nu_{s,init}$ and $\nu_{s,final}$, and one
would generally expect $\sigma^{ref}_{ab} \ne 0$.

The point of these definitions is the following.  Imagine a NS with
strained crust and spin $\nu_s$. To determine this NS's precession
frequency, we want to know what residual oblateness $\Delta I_d$ it
would have if it were spun down to zero frequency. We can imagine
accomplishing the spindown in two steps. First, adjust the spin to
$\nu_{s,ref}$. At this spin, the crust has some strain
$\sigma^{ref}_{ab}$, but $\sigma^{ref}_{ab}$ has no net effect on the
NS's oblateness. Second, we spin the star down from $\nu_{s,ref}$ to
zero angular velocity. We define $\sigma^{sd}_{ab}$ to be the
extra strain that this spindown induces; i.e., 
$\sigma^{sd}_{ab} \equiv \sigma_{ab}
- \sigma^{ref}_{ab}$, where $\sigma_{ab}$ is the strain in the NS
when it is nonrotating.
Conceptually, we can consider both
$\sigma^{ref}_{ab}$ and $\sigma^{sd}_{ab}$ as independent, small 
(i.e., linear) 
perturbations, and work to first order in both.
To this order, $\sigma^{sd}_{ab}$ is just the strain induced when
spinning down a relaxed NS from $\nu_{s,ref}$ to zero spin, which we
solved for in \S 4. Also to this order, 
the oblateness $\Delta I_d$  caused by $\sigma_{ab}$ is the sum 
of (i) the oblateness
due to $\sigma^{sd}_{ab}$ and (ii) the oblateness due to 
$\sigma^{ref}_{ab}$, but (ii) vanishes, by definition. 
That is, $\Delta I_d$ is just the oblateness
induced by the spindown part of the strain field,
Thus, generalizing Eq.~(1), we can write \be\label{DId_gen} \Delta I_d
= b\, \Delta I_{\Omega}(\nu_{s,ref}) \, , \ee
\noindent
where $\Delta I_{\Omega}(\nu_{s,ref}) \approx 0.3\, I_0 \bigl(\nu_{s,ref}/{
\rm kHz}\bigr)^2$ is the oblateness of a fluid star with spin
$\nu_{s,ref}$, and $b$ is the same coefficient (i.e., same numerical
value) we calculated in \S~\ref{sec:crust-deformation}.

\section{The precession frequency of PSR 1828-11}\label{sec:1828}
In \S 2 we showed that the measured precession period of 
PSR 1828-11 implies that for this NS,
\be\label{eq:DIdmeas}
{\Delta I_{d}\over I_C}|_{1828} = 9.2\times 10^{-9} \left ({511\mbox{ d}\over
P_p}\right ) \, .
\ee
For a NS with relaxed crust (i.e., relaxed except for the 
stresses induced by precessional motion itself), the predicted value is
\be\label{eq:predict}
{\Delta I_d\over I_C} = b \frac{\Delta I_\Omega}{I_0} \left
(\frac{I_C}{I_0}\right )^{-1} \, .
\ee
From the results of \S 4, we estimate (for our Newtonian NS models):
\begin{eqnarray}
\Delta I_{\Omega}/I_0 &\approx& 
0.3 \left (\frac{\nu_s}{\rm kHz}\right )^2\,\left ({M\over 1.4
M_\odot}\right )^{-1}\left ({R\over 12\mbox{ km}}\right)^3 \, , \label{IOMscale} \\
I_C/I_0 &\approx& 0.01 \left ({M\over 1.4 M_\odot}\right )^{-2}\left
({R\over 12\mbox{ km}}\right )^4 \, , 
\label{ICscale} \\
b &\approx& 2\times 10^{-7} \left ({M\over 1.4 M_\odot}\right
)^{-3}\left ({R\over 12\mbox{ km}}\right)^5 \, . 
\label{bscale} 
\end{eqnarray}
The scaling with
NS mass and radius are estimated by taking (i) $\Delta I_\Omega/I_0
\propto MR^2/(M^2/R) = R^3/M$ (i.e., to the ratio of the NS's kinetic
and potential energies), (ii) $I_C/I_0 \propto M_C/M \propto
\rho_{bot} R^2\Delta R\,/M \propto R^2 p_{bot}/(g M) \propto R^4/M^2$,
and (iii) $b \propto \mu_{ave} V_C/(M^2/R) \propto R^2\Delta R/(M^2/R)
\propto R^2 (p_{bot}/(g \rho_{bot}))/(M^2/R) \propto R^5/M^3$.  Here
$M_C$ is the mass of the crust, $\Delta R$ is its thickness, $V_C$ its
volume, $\rho_{bot}$ and $p_{bot}$ are the density and pressure at the
bottom of the crust, $\mu_{ave}$ is a volume-weighted average of the crustal
shear modulus, and $g \equiv GM/R^2$.  
Taking $\nu_s = 2.5\,$Hz, we
estimate for PSR 1828-11: 
\be
\label{eq:DIdpredict} 
{\Delta I_{d}\over I_C}|_{\rm relaxed} \approx 3.8 \times
10^{-11}\,\left ({I_C/I_0}\over
10^{-2}\right ) \left ({M\over 1.4 M_\odot}\right)^{-2}\left({R\over
12\mbox{ km}}\right)^4 \, .  \ee Note that the measured value,
Eq.~(\ref{eq:DIdmeas}), is $\sim 250$ times larger than our
relaxed-crust estimate, for our fiducial NS parameters and precession
period.  And note that if some substantial fraction of the core is
dynamically coupled to the crust (so $I_C >> 0.01 I_0$), then the
discrepancy only gets worse. Even if we assume
that PSR 1828's precession period is really $\sim 1000$ days and take
rather extreme values for the NS mass and radius, $M = 1.0 M_\odot$
and $R = 20$ km, we are still left with a factor $\sim 8$
discrepancy. 
We conclude that either our basic
picture is wrong (i.e., either PSR 1828-11 is not actually precessing,
or that some other mechanism {\it besides} crustal shear stress is
responsible for its large $\Delta I_d$), or that the crust is {\it
not} relaxed. Here we adopt the latter explanation, and pursue its
implications.

The 511-d precession period for PSR B1828-11 implies that the star's
reference spin is $40$ Hz. Subsequent spin down to its present spin of
2.5 Hz without significant structural relaxation (through quakes or
plastic flow) would strain the crust and give the inferred $\Delta
I_{d}/I_C=9.2\times 10^{-9}$. What are the current strain levels in
the crust of 1828-11 in this scenario?  Compared to $40$ Hz, its
current spin of $2.5$ Hz is very slow. If one continued slowing PSR
1828-11 down to zero angular velocity, the spindown-induced crustal
stresses would increase fractionally by only $0.4\%$. So (neglecting
that $0.4\%$), the current strains are the sum of 
a) the reference strain $\sigma^{ref}_{ab}$ that this NS would
still have if it were spun up to $\nu_s = 40\,$Hz and 
b) the spindown strains $\sigma^{sd}_{ab}$ induced
by spinning a relaxed NS down from $\nu_s = 40\,$Hz to $\nu_s = 0$.
While $\sigma^{ref}_{ab}$ is practically unknowable, the
spindown-induced strains $\sigma^{sd}_{ab}$ are those we found in
\S 4.

We parametrize the strength of the spindown strain $\sigma^{sd}_{ab}$ by
the scalar quantity $\bar{\sigma}^{sd}$, defined by

\begin{equation}\label{eq:bar_sigma_sd}
\bar{\sigma}^{sd} \equiv \biggl(\frac{1}{2}\sigma^{sd}_{ab}\sigma^{sd,ab}\biggr)^{1/2} \, ,
\end{equation}
\noindent 
and we define the $\mu$-weighted-average of the crustal spindown strain,
$\bar\sigma_{\rm ave}$,  by 
\be \label{eq:ave}
\bar\sigma^{sd}_{\rm ave} \equiv \frac{\int{\mu\, \bar\sigma\, dV}}{\int{\mu\, dV}} \, .
\ee
We have computed  $\bar\sigma^{sd}_{\rm ave}$ for our five EOS; the
results are in Table~\ref{table:eos-summary}. They all give roughly the same result: 
$\bar\sigma^{sd}_{ave} \approx 0.5$ for $R\Omega/c = 1$.
Scaling to the $40\,$Hz reference angular velocity of PSR1828-11 yields
\be\label{big_result}
\bar\sigma^{sd}_{\rm ave} = 5\times 10^{-5} 
\left(\frac{P_s}{0.4\, {\rm s}}\right)
\left({P_p\over 511\mbox{ d}}\right)^{-1}\left({I_C/I_0\over 0.01}\right) 
\left ({b\over 2\times 10^{-7}}\right )^{-1} 
\, .
\ee

Again, the total strain $\sigma_{ab}$ is the sum of
$\sigma^{sd}_{ab}$ and $\sigma^{ref}_{ab}$, where $\sigma^{ref}_{ab}$
is determined by the detailed evolution of the crust, through
structural adjustments, 
and so without knowing that history we can say
nothing definitive about $\sigma^{ref}_{ab}$ (except that, by
definition, these strains have no net effect on the NS's quadrupole
moment). However, since $\sigma^{ref}_{ab}$ will contain contributions
from all harmonics (i.e., all $Y_{lm}$), while $\sigma^{sd}_{ab}$ is
wholly $l=2, m=0$, it seems quite likely that the NS's average total
strain is larger than the average spindown strain. 
{\emph If} this is true, the average value of spindown portion of the
strain places a lower limit on the crustal breaking strain:
$\bar \sigma_{max} \agt 5 \times 10^{-5} (511 {\rm d}/P_p)$.
In terrestrial solids, strains of order $5 \times 10^{-5}$ are rather
easy to maintain, so it seems likely that the NS crust is strong
enough to sustain the $\Delta I_d$ implied by the precession frequency.
If $I_C/I_0$ is of order unity,
the implied average strain would be 
$\bar \sigma_{ave} \agt 5 \times 10^{-3} 
(511 {\rm d}/P_p)$, which is still plausibly below the
crust's breaking strain.
We conclude that the neutron star crust is likely
strong enough to deform PSR B1828-11 to the extent that it will 
precess with a period of $\sim$ 500 d.

The spindown strain field suggests that the evolutionary picture of
crustal strain might be as follows. Suppose the star is born relaxed
and rapidly-spinning. As the star spins down the strain becomes
largest first in the upper crust (see
Fig. \ref{fig:sigma-2d}). Cracking begins in the upper crust when the
yield strain is reached there; however, because relatively little strain
energy can be stored in this region of small shear modulus, these
small quakes would not be expect to relax the strain that has developed
deeper in the crust. When the yield strain is reached deeper,
larger (\ie, more energetic) quakes occurs there. Eventually an
equilibrium is reached in which the strain field is just sub-critical
throughout the crust. If the Earth is any guide, then numerous small 
quakes should occur often, while large quakes
occur rarely, but release most of the accumulated strain energy. These
large events could excite the precession. We will explore this
scenario further in a future publication.

In this picture, one expects that in the upper crust, $\sigma^{ref}_{ab}$
nearly cancels $\sigma^{sd}_{ab}$ (so that the sum is below the
breaking strain), while near the bottom of the crust, where
spindown strain in lower, the actual strain level $\bar \sigma$ is
probably greater than $\bar \sigma^{sd}$. E.g., if the bottom of the crust has
not significantly relaxed since the NS's spin was $60$ Hz, then the 
total strain field near the bottom would presumably 
resemble $\sigma^{sd}_{ab}$ for
spindown from $\nu_{s,ref} = 60\,$Hz.

\section{Other candidate precessing pulsars}
So far we have focused our discussion on PSR 1828-11; we now apply our
results to other pulsars showing modulations that may be due to
precession. Here we discuss the Crab, Vela, PSRs B1642-03, B2217+47,
B0959-54, and the possible remnant in SN 1987A.  The strength of
evidence for precession of these pulsars varies from more or less
convincing to marginal; only one of these candidates (PSR
B1642-03) shows strongly periodic {\it and\/} correlated changes in both
pulse shape and phase similar to those observed in PSR 1828-11.  In
this section we merely summarize all published claims of precession,
without attempting to assess their validity. 
Our discussion will be brief, since most these cases have recently
been reviewed in depth by Jones \& Anderson (2001). Note however that
our conclusions are often different from Jones \& Andersson (2001),
because our more accurate value of $b$ is nearly two orders of magnitude
smaller. Our results are summarized in Table 2.

Before proceeding, we note that two of the precession candidates
that we discuss, the Crab and Vela pulsars, suffer glitches in spin
rate. Glitches are thought to represent angular momentum transfer to
the crust from the interior superfluid as the array of superfluid
vortices, usually pinned to nuclei of the crust or magnetic flux tubes in the
core, undergoes a sudden expansion (see, {\sl e.g.}, Anderson \& Itoh
1975; Link \& Epstein 1996; Ruderman, Zhu \& Chen 1998). Even a small amount of
vortex pinning is inconsistent with long-period precession (Shaham
1977). For the sake of this discussion, we ignore this objection. We
refer the reader to Link \& Cutler (2002) for a discussion of how
precession could unpin vortices that are intially pinned to the crust.

\subsection{PSR B1642-03}

PSR B1642-03 has  $\nu_s = 2.5\,$ Hz (the same spin rate as PSR 1828-11), and
shows a $10^3\,$ day periodicity in pulse shape, with modulation amplitude 
$\approx 0.05$ (Cordes 1993). This period is $120$ times shorter
than the free-precession period, assuming a relaxed crust.
If shear stresses are responsible for PSR B1642-03's $\Delta I_d$, then
its reference angular velocity is  
$\sim 27\,$Hz. 

\subsection{PSR B0531+21 (the Crab pulsar)}

The Crab pulsar has $P_s = 0.0331\,$s; Lyne et al. (1988) observed a phase
residual with period $20$ months, which Jones (1988) interpreted as
evidence  for precession. Thus $P_s/P_p = 6.4\times 10^{-10}$, while for
our fiducial NS, we predict $\Delta I_d/I_C \approx 5.4\times 10^{-9}$
This factor $\sim 8$ discrepancy is not large considering the 
uncertainties in the NS EOS, crust thickness, and the mass and radius of
the Crab; we conclude that a 20-month precession period is 
not in significant disagreement with our theoretical estimate.
This conclusion is contrary to that of Jones \& Anderson (2001), who
concluded that the discrepancy was a factor $\sim 700$ instead of $\sim 8$.

\subsection{PSR 2217+47}
PSR 2217+47 has $P_s=0.538$~s, and was originally thought to have a
single-component pulse profile.  However, Suleymanova \& Shitov (1994)
reported the discovery of a weaker second component that varies with a
period of $\sim6-8$~years. Additionally, the braking index of 2217+47
has changed significantly between the original observation epoch
(1974-1984) and the subsequent observations (Shabanova 1990).
Suleymanova \& Shitov (1994) interpret the
change in the spindown rate as evidence for free precession with
period at least comparable to the baseline of observations (i.e.,
$P_p \gtrsim20$~years), and attribute the more rapid pulse shape
variation to a patchy structure of the emission beam.  Taking this
interpretation at face value, the reference angular velocity of the
crust is $\nu_{s,ref} \lesssim12$~Hz (compared to the current
$\nu_s=1.86$~Hz) and  $\bar\sigma^{sd}_{\rm ave} \lesssim 4\times10^{-6}$.

\subsection{PSR B0959-54}

D'Alessandro \& McCulloch (1997) show that the odd moments of the
pulse profile of B0959-54 are negatively correlated with the
variations in the timing residuals.  The timing residuals appear
quasi-periodic, and so they interpret this observation as evidence for
free precession with $P_p \gtrsim2500$~days (i.e., greater than or
equal to the period of the timing residuals).  This constraint on
$P_p$ implies the crust's reference angular velocity is $\nu_{s,ref}
\lesssim33$~Hz (compared to spin frequency $ \nu_s = 0.696$~Hz), and
$\bar\sigma^{sd}_{\rm ave} \lesssim 3\times 10^{-5}$.  

\subsection{PSR B0833-45 (the Vela pulsar)}

The  Vela pulsar has $P_s = 0.089\,$s; Deshpande \& McCulloch (1996)
detected intensity variations with a $165$-d period, which they
interpreted as possible evidence of precession. 
Thus $P_s/P_p = 6.2\times 10^{-9}$, while for
our fiducial NS, we predict $\Delta I_d/I_C \approx 7\times 10^{-10}$.
The factor $\sim 9$ discrepancy could be accounted for by Vela 
having a rather small mass and correspondingly large radius 
(e.g., $M \approx 1.1 M_\odot$ and $R \approx 18$ km), 
or by Vela having a crust whose reference angular velocity is
$\nu_{s,ref} \approx 32\,$ Hz (i.e., $3$ times the current spin). 
The latter would entail an average crustal spindown strain of 
$\bar\sigma^{sd}_{ave} \sim 3 \times 10^{-5}$.

\subsection{Remnant in SN1987A}

Middleditch et al.~(2001) have suggested 
that there is a precessing pulsar in the remnant of
SN 1987A. They present evidence for the existence of a pulsar (now faded from
view) with spin period $P_s = 2.14\,$ms, and evidence for modulation (which
they interpret as precession) on a timescale $P_p \sim 10^3\,$s, requiring
$\Delta I_d/I_C \approx 2\times 10^{-6}$. Though failure to observe
a pulsar in SN1987A since 1993 brings the existence of a pulsar into
question, we apply our results assuming the pulsar exists.
For our fiducial NS parameters,
we would expect $\Delta I_d/I_C \approx 1.3\times 10^{-6}$. 
The factor $\sim 1.5$ difference is small compared to 
uncertainties in the NS EOS, mass, etc. 
We conclude (in contrast to Jones \& Andersson 2001) 
that the $10^{3}$s precession timescale is quite reasonable, 
theoretically.

\subsection{Discussion}

Figure~\ref{fig:prec-limit} summarizes the observational claims of
precession presented in this section. In this figure, we plot the
observed spin periods of neutron stars against their precession
claimed periods (or lower limits on $P_p$ for PSRs B2217+47 and
B0959-54). What limits can we place on the precession periods using
the interpretation adopted in this paper, i.e., that the precession is
due to a non-spherical shape of the crust? First, $P_p=P_s(I_C/\Delta
I_d)=P_s (I_C/b\Delta I_\Omega)\propto \nu_{s,ref}^{-2}$ (for given 
$\nu_s$).  
Dashed lines in Figure~\ref{fig:prec-limit} show the above relation
for $1/\nu_{s,ref}=1$~ms and 20~ms, respectively. A pulsar lying
below the 1~ms line would have a reference spin frequency faster
than 1~kHz. Note that none of the pulsars with claims of precession
require uncomfortably large $\nu_{s,ref}$, i.e., nowhere close to
breakup spin of $\sim 1.5$~kHz
for our fiducial mass and radius. (See
Cook et al. 1994 for an exhaustive summary of breakup frequencies for
various equations of state.) 
Second, if the crust of the neutron star
is relaxed at its current spin frequency, then 
$P_s=1/\nu_{s,ref}$ and $P_p\propto\nu_{s,ref}^{-3}$.  This
relation is shown by the solid line in Figure~\ref{fig:prec-limit}.
Since we presume that the NS had a relaxed crust in the past, and has
since then spun down, theoretically feasible precession
candidates must lie to the right of the solid line. All precession
candidates except for the Crab pulsar satisfy this constraint.  
The Crab's claimed precession frequency can also be explained in 
terms of its crustal rigidity, if one
assumes that its mass and radius are somewhat different 
from our fiducial values (e.g., $M = 1.8 M_{\odot}$ and $R = 8.0\,$km).

\section{Summary}

A relaxed, self-gravitating object with a rigid crust has a portion of
its spin-induced bulge that cannot follow changes in the direction of
the solid's rotation vector. Such an object can precess, at a
frequency that is determined by the material properties of the crust
and the strength of gravity, whose relative strengths are determined
by the rigidity parameter $b$. In this paper showed that a NS's
rigidity parameter $b$ is a factor $\sim 40$ times smaller than the
estimate of Baym \& Pines (1971; see Eq.~\ref{eq:bBP}).  Using this
result, we showed that the precession frequency $\nu_p$ for PSR
1828-11 is $\sim 250 (511 {\rm d}/P_p)$ times faster than expected for
a NS with a relaxed crust, indicating that its crust is significantly
strained. The strain would naturally arise from the secular
spin down of the star. Large quakes that partially relax the
accumulated strain energy could also excite precession. Applying our
results to PSR 1828-11 implies the average value of spindown strain in
the crust is is $\bar\sigma^{sd}_{ave} \sim 5\times 10^{-5}$ if the
core and crust are effectively decoupled, and larger by a factor of
$\simeq 100$ if the crust and core are dynamically coupled over a spin
period. It is not unreasonable to expect the NS crust to be able to
sustain such modest strain values.  Applying our model of a relaxed
crust to other precession candidates, we found that only SN1987A and
the Crab pulsar are consistent with the hypothesis of a relaxed crust.
The other candidates can be reasonably explained as having strained
crusts, though unfortunately this explanation has no predictive power;
the one free parameter (the reference spin) is adjusted to fit the
precession timescale.

We conclude here by
mentioning one other application of our result for $b$. 
Cutler (2002) has shown that for rapidly rotating NS's
with relaxed crusts and a strong, interior toroidal magnetic field
$B_t$, the prolate distortion of the star induced by $B_t$ dominates
over the oblateness frozen into the crust for $B_t > 3.4 \times
10^{12}{\rm G} (\nu_s/300{\rm Hz})^2$.  
In this case, dissipation
tends to drive the magnetic symmetry axis orthogonal to the spin
direction, and the NS becomes a potent gravitational wave emitter.
Had one used the Baym-Pines value
for $b$, one would have arrived at a substantially larger requirement 
on the toroidal field (in order for it to dominate over crustal
rigidity):  
$B_t > 1.4 \times 10^{14}{\rm G} (\nu_s/300{\rm Hz})^2$.

\acknowledgements
C.C.'s work was supported in part by NASA Grant NAG5-4093,
B.L.'s work was supported in part by NSF Grant AST 00-98728, and
G.U.'s work was supported by a Lee A. 
DuBridge postdoctoral fellowship at Caltech.  

\appendix

\section{Approximate $b$ from integral of crustal stresses}
\label{appendix:check}

Our solution for the NS's rigidity parameter $b$ in
\S~\ref{sec:crust-deformation}-\ref{sec:results} involved calculating
the exterior $l=2$ piece of the gravitational potential $\delta \phi$
for two stars spun up to angular value $\Omega$: i) a star that is
completely fluid, and, ii) a star with relaxed, spherical crust that
is otherwise identical to the first one.  The relative difference in
the exterior $\delta \phi$ {\it is} $b$.  
Our result (Table 1) is that $b \sim 2\times 10^{-7}$, so an
accurate calculation of $b$ by the above subtraction method requires
solving for $\delta \phi$ to $\sim 9$ decimal places. Here, as an
additional check on our work, we estimate $b$ in a way that does not
require such high accuracy from our solutions.

UCB derived an identity relating a 
NS's mass multipole moment $Q_{22}$ to an integral of stresses
in the crust. The same calculation can be repeated for the $Q_{20}$
multipole moment. Defining $Q_{20}$ by

\be
Q_{20} \equiv \int {\delta \rho(r,\theta,\phi) Y_{20}(\theta,\phi) dV}
\ee

\noindent we obtain 

\begin{eqnarray}\label{eq:qform}
Q_{20} = &-\bigl(1-F\bigr)^{-1}&\int\frac{r^3}{g}\biggl[\, 
        \frac{3}{2}\left(4 - \tilde U\right) t_{rr}  
+\frac{1}{3} \left(6 - \tilde U\right) t_{\Lambda} \\ \nonumber
&+&\sqrt{\frac{3}{2}} \left(8 - 3\tilde U + \frac{1}{3} \tilde U^2 
-\frac{1}{3} r\, \frac{d\tilde U}{dr}\right)t_{r\perp}\, \biggr]\,dr,
\end{eqnarray}
\noindent where $g(r) = G m(r)/r^2$, 
$\tilde{U}(r) \equiv 4\pi\rho r^3/m(r)$ (note $\tilde{U}(r) << 1$ in the crust), 
and the stress components of the
stress tensor $t_{ab}$ are defined by
\begin{eqnarray}\label{eq:max-tab}
t_{ab} &=& t_{rr}(r) Y_{lm} (\hat r_a \hat r_b -\frac{1}{2}  e_{ab})
\\ \nonumber
&+& t_{r\perp}(r) f_{ab} 
+ t_{\Lambda}(r) (\Lambda_{ab} + \frac{1}{2} e_{ab}) ,
\end{eqnarray}
\noindent where the tensors $e_{ab}$, $f_{ab}$, and $\Lambda_{ab}$ were
defined in Eq.~(\ref{eq:dtab-term-abbreviations}). 

The factor $(1-F)^{-1}$ in Eq.~(A2) requires some explanation.
UCB derived Eq.~(A2), without that factor, within the {\em Cowling 
approximation}, in which the self-gravity of the perturbation is
neglected. So $(1-F)^{-1}$ is a correction factor that accounts 
the effect of the perturbation's self-gravity. UCB derived an approximate
expression for F (UCB, Eq.~72), leading to the estimate that 
$F \approx 0.2-0.5$, depending 
on the particular NS model. 

We have calculated the right-hand side of Eq.~(\ref{eq:qform}) for the
stresses obtained by solving our problem (ii), with the source term
$(R\Omega/c)^2$ set equal to $-1$.  The result is the residual
$Q_{20}$ for a NS whose crust was relaxed at spin $\Omega = c/R$, and
which was then spun down to $\nu_s =0$.  We find $Q_{20}|_{\rm
residual} = (1-F)^{-1} 4.24 \times 10^{38}$g-cm$^2$ for our fiducial
$1.4 M_\odot$ NS, constructed with the AU EOS.  


  The rigidity parameter $b$ is the residual $Q_{20}$ of the non-rotating
NS, divided
by the $Q_{20}$ of the rotating model. The latter is given by
\be
Q_{20}|_{\rm rotating} = \frac{5 c^2 R^3}{4\pi G} z_5|_{r=R}, 
\ee
where $z_5 = \delta \phi/c^2$. For $\Omega$ set equal to $c/R$, we
find $Q_{20}|_{\rm rotating} = 2.8 \times 10^{45}$g-cm$^2$ 
for the same fiducial model.
Thus integrating the stresses in the crust yields
$b = (1-F)^{-1} \times 1.5 \times 10^{-7}$, or, for the $F$ in the range
$0.2-0.5$, $b \approx 1.9-3.0 \times  10^{-7}$. This is in excellent
agreement with the value $b = 2.47 \times 10^{-7}$ in Table 1.

\section{Analytic solution for $b$ for homogeneous sphere 
with thin crust}
\label{appendix:analytic}

A classic problem solved by Lord Kelvin was the determination of  
$b \equiv \Delta I_d/\Delta I_{\Omega}$ for the
case of a constant-$\rho$, constant-$\mu$ sphere; the result
can be written as $b =\frac{57}{10}\mu V/|E_g|$, where $V$ is the star's
volume and $E_g$ is its binding energy. (For the derivation, see the
classic treatise by Love 1944.)
In Fig.~9 we plot $d\,E_{strain}/dr$ for this case. We see 
that the strain energy is concentrated toward the center of the star, which
already suggests that this model problem will not be a reliable guide for 
estimating the rigidity of a realistic NS (where all the shear stresses
are near the surface.)

Here we solve the corresponding
problem for a star with a thin crust. That is, we consider a constant-$\rho$ 
star that consists of two pieces: a fluid interior (where
the shear modulus $\mu$ is zero), plus a thin crust where $\mu$ is
constant. Our case is clearly much closer to a realistic NS, where
the ratio (crust thickness)/(NS radius) is $\sim 1/20$. 
Most of the analysis we require has already been carried out by
Franco, Link \& Epstein (2001), who solved for the displacements and
strain build-up that occur when such a star spins down. 
Franco, Link \& Epstein  (2001) 
work within the Cowling approximation--i.e., they neglect the 
gravitational perturbation $\delta \phi$ induced by crustal
distortions--so for convenience in this Appendix we neglect 
$\delta \phi$ as well. As explained in Appendix A, the
Cowling approximation underestimates $b$ by a factor $\approx 5/3$.

Franco, Link \& Epstein (2001) 
did not restrict themselves to the case of a {\it thin} crust, but we
do so, for convenience. 
Rather than repeat their derivation, we shall simply
quote their formulae; e.g., we shall refer to their 39th numbered
equation as (F39). Franco et al.
define $R^\prime$ and $R$ to be
the radii at the bottom and top of the crust, respectively.
We define $\Delta R \equiv R - R^{\prime}$ and neglect terms that
are of quadratic or higher order in $\Delta R$. We just highlight 
the basic steps and refer the reader to  Franco et al.(2001) for 
further details. 

We will show
that for the thin-crust case, and within the Cowling approximation,
\begin{equation}
\label{thin}
b =\frac{12}{11}{\mu V_c\over |E_g|}
\end{equation}
\noindent  where $V_c$ is 
the volume of the crust and $E_g$ is still the binding energy of the 
entire star.

When the star's spin-squared changes by $\delta(\Omega^2)$, the crust
undergoes a displacement $\vec u(r,\theta)$, whose radial piece
can be shown to have the form $u_r(r,\theta) = f(r) P_2(\theta)$,
where $P_2$ is the second Legendre polynomial, and where $f(r=R)$ 
has the form:
\begin{equation}
\label{fR}
f(R) = {5\over 6}R {{\delta(\Omega^2)}\over{v^2_k}}\bigl(1 - b\bigr) \, .
\end{equation}
Here $v^2_k \equiv GM/R$ and $M$ is the star's mass.
Eq.~(F31) shows that the Eulerian change in the ($l=2$ piece of the) 
gravitational potential exterior to the star is $\propto f(R)$, which
makes it clear that the $b$ on the right-hand side of Eq.~(\ref{fR}) must also
equal $\Delta I_d/\Delta I_{\Omega}$. Expanding (F23) and (F24) to 
linear order in $\Delta R$ and subsituting into (F19), we find 
\begin{eqnarray}
f(R) = {1\over 3} AR^3 - {3\over{16}}BR^{-2} \label{fR2a} \\
f'(R) = {1\over 4} AR^2 + {3\over 4}BR^{-3} \label{fR2b} \\
f'(R-\Delta R) = f'(R) - \Delta R [-2AR -{3\over 4}BR^{-4}] \label{fR2c}
\end{eqnarray}
where $f' \equiv df/dr$ and $A$ and $B$ are coefficients (to be solved for) 
appearing in the expansion of $f(r)$; see F19. Expanding F38 to linear
order in $\Delta R$ and comparing with Eq.~(\ref{fR2c}), we can solve
for $B$ in terms of $A$: 
\begin{equation}\label{BofA}
BR^{-3} = {-2\over 3}AR^2 + {2\over 5}AR\,\Delta R \,.
\end{equation}
Plugging (\ref{BofA}) into Eqs.~(\ref{fR2a})-(\ref{fR2b}) yields
\begin{eqnarray}
f(R) =   {{11}\over {24}}AR^3 - {3\over {40}}AR^2\,\Delta R  \label{fR3a}\\
f'(R) =  -{1\over 6}AR^2 + {3\over {10}}AR\,\Delta R  \label{fR3b} \, , 
\end{eqnarray}
and plugging Eqs.~(\ref{fR3a})-(\ref{fR3b})into (F33) 
yields
\begin{equation}\label{fR4}
f(R) = {5\over 6}R{{\delta(\Omega^2)}\over{v^2_k}}\bigl(1- {{60}\over{11}}{{\Delta R}\over R}{{c^2_t}\over {v^2_k}}) \, .
\end{equation}
\noindent where $v^2_k \equiv \mu/\rho$.  The second term in
parentheses on the righ-hand side of Eq.~(\ref{fR4}) is $b$, and can
be re-expressed as
\begin{equation}
\label{thin2}
b =\frac{12}{11}\mu V_c/|E_g| \, .
\end{equation}

Note that the coefficient $12/11$ in Eq.~(B10) is smaller than the
$57/11$ in the Baym-Pines estimate, Eq.~(\ref{eq:bBP}), by
$209/40 \approx 5.23$. So from our uniform, thin-crust model, we 
see that the Baym-Pines expression significantly overestimates
the rigidity of a NS.

If we were to simply replace the $57/11$ in the Baym-Pines estimate
by $12/11$, we would {\it still} overestimate the rigidity of a realistic
NS by a factor $\sim 8$. To understand this factor, it is instructive to
to evaluate $Q_{20}|_{\rm residual}$ for the uniform, thin-crust case
using our integral expression Eq.~(A2), and compare with the realistic
case. For the uniform, thin-crust case, $\tilde U = 3$, and
using Eq.~(76) in UCB we find that the correction factor
$(1-F)^{-1}$ is exactly $5/2$. The results in Franco et al.(2001) and 
a few pages of algebra suffice to show that, to leading order in
$\Delta R$,
\be\label{exact-sig}
\sigma_{rr} = \frac{10}{33}\sqrt{\frac{4\pi}{5}}\frac{R^3 \Omega^2}{GM} \, , \ \ \sigma_{\Lambda} = \frac{10}{33}\sqrt{\frac{4\pi}{5}}\frac{R^3 \Omega^2}{GM} \, , \ \ \sigma_{r\perp} = 0 \, ,
\ee
\noindent for the case of star spun down from $\Omega$ to zero
angular velocity. Also, for a uniform star rotating at angular velocity 
$\Omega$, 
\be\label{q20}
Q_{20} = \frac{5}{6}\sqrt{\frac{4\pi}{5}}\rho R^5 \frac{R^3 \Omega^2}{GM} \, .
\ee
Using these results and Eq.~(A2), we find that the term $\propto
\sigma_{rr}$ in Eq. (A2) contributes $(-\frac{1}{2})\frac{12}{11}\mu
V_c/|E_g|$ to $b$, while the term $\propto \sigma_{\Lambda}$
contributes $(\frac{3}{2})\frac{12}{11}\mu V_c/|E_g|$.

Comparing these intermediate results to those for the realistic NS case
(with compressible matter and a steep density gradient), we see that in the
realistic case, there is much more cancellation within the integral
for $Q_{20}$. 
As shown in Fig.~10, in the realistic case both $\sigma_{rr}$ and
$\sigma_{\Lambda}$ switch signs at different depths inside the crust
(so that the contributions to $b$ from different layers tend to cancel
each other, unlike in the uniform, thin-crust case). Also, the
$\sigma_{rr}$ and $\sigma_{\Lambda}$ contributions are clearly
much closer in magnitude (though still of opposite sign) in the realistic
case, and so cancel each other much more nearly.

\clearpage

\begin{figure}
\epsfig{file=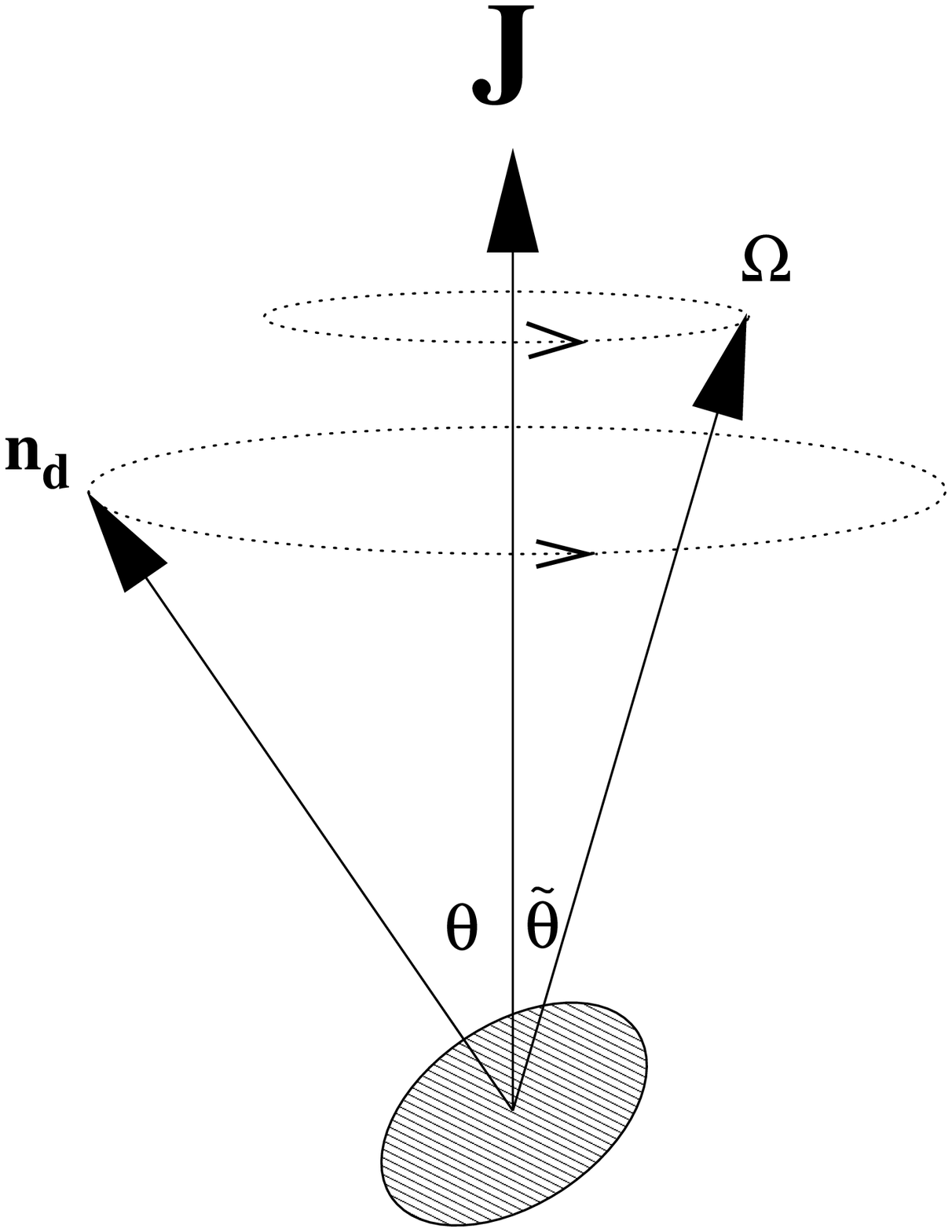}
\caption{\label{fig:angles} Definitions of the angles $\theta$ and
$\tilde\theta$ in relation to the angular momentum vector $J^a$, the body
axis $n_d^a$ and the spin axis $\hat\Omega^a$.}
\end{figure}

\begin{figure}
\epsfig{file=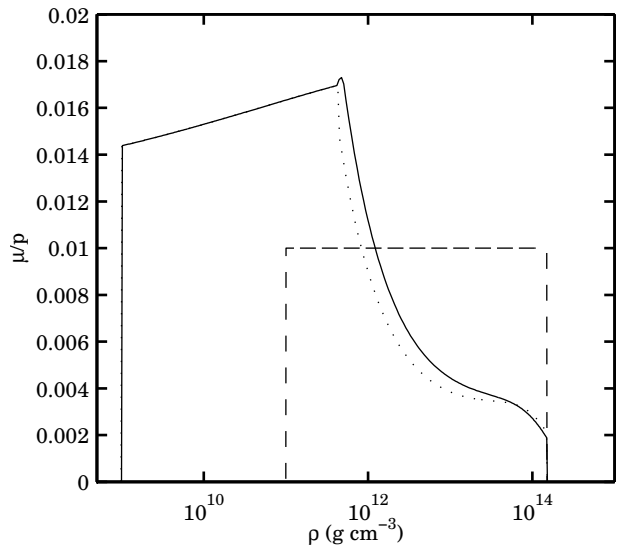}
\caption{\label{fig:mu} The ratio of the shear modulus to the pressure
as a function of density for the NS models used in this paper: model
AU (solid line), WS (dotted line), n1poly (dashed line).}
\end{figure}

\begin{figure}
\epsfig{file=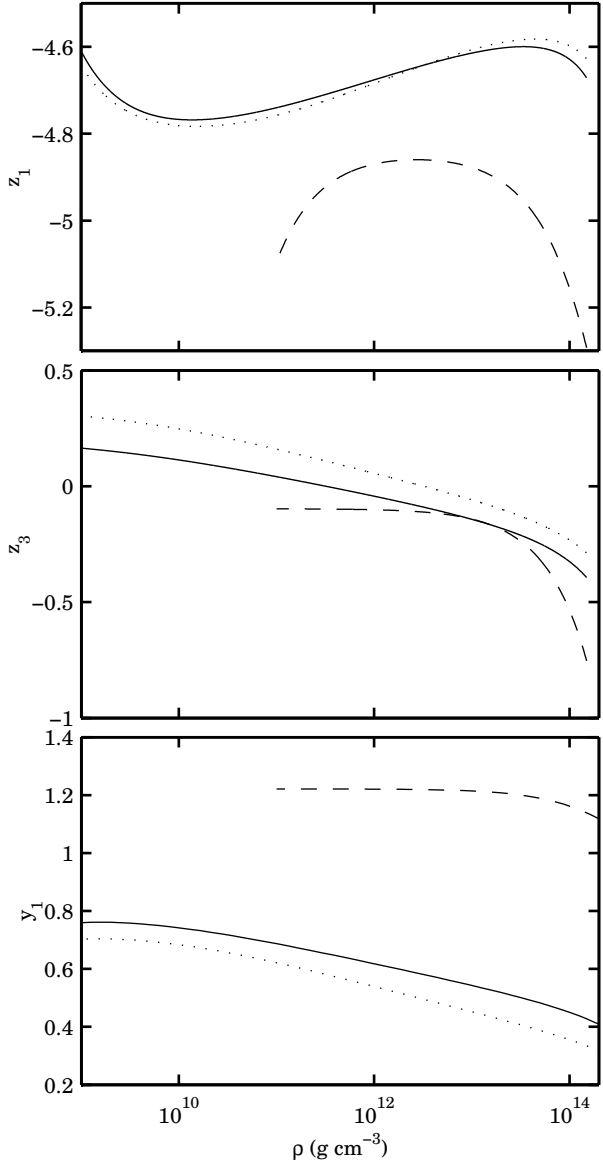}
\caption{\label{fig:z1} Crustal displacements for a relaxed, 
nonrotating NS that is 
spun up to $\Omega = c/R$. The different curves represent 
different equations of
state: model AU (solid lines), WS (dotted lines), and
n1poly (dashed lines).   
For $\nu_{s,ref}=40$~Hz,
the values in this plot must be multiplied by $6.5\times 10^{-5}$.
Top panel: $l=2$ radial displacement $z_1$; middle panel: $l=2$
transverse displacement $z_3$, bottom panel: $l=0$ radial displacement
$y_1$.}
\end{figure}

\begin{figure}
\epsfig{file=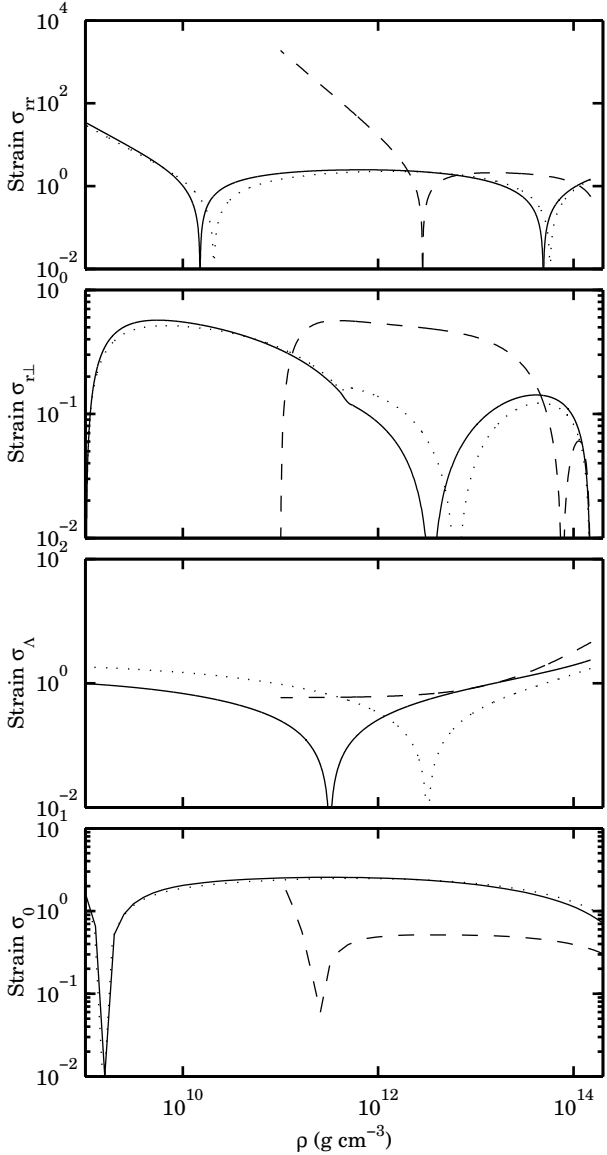}
\caption{\label{fig:sigma-all} Radial dependence of the 
(spindown portion of the) 
strain tensor
for model AU (solid line), WS (dotted line), and n1poly (dashed
line), normalized to $\Omega = c/R$.  
For $\nu_{s,ref}=40$~Hz, the values in this plot must be multiplied by $6.5\times10^{-5}$.
Top panel: $|\sigma_{rr}|$, second panel: $|\sigma_{r\perp}|$, third
panel: $|\sigma_\Lambda|$, bottom panel: $|\sigma_0|$.
The signs of the components are as follows.
At the bottom of the crust, for all 3 EOS: $\sigma_{r\perp} > 0$,
$\sigma_{\Lambda} < 0$, and $\sigma_0 > 0$. $\sigma_{rr} > 0$ at the
bottom for the AU and  WS EOS, but $\sigma_{rr} < 0$ there for
the n1poly model. The sharp dips in the figures are zero-crossings.}
\end{figure}

\begin{figure}
\epsfig{file=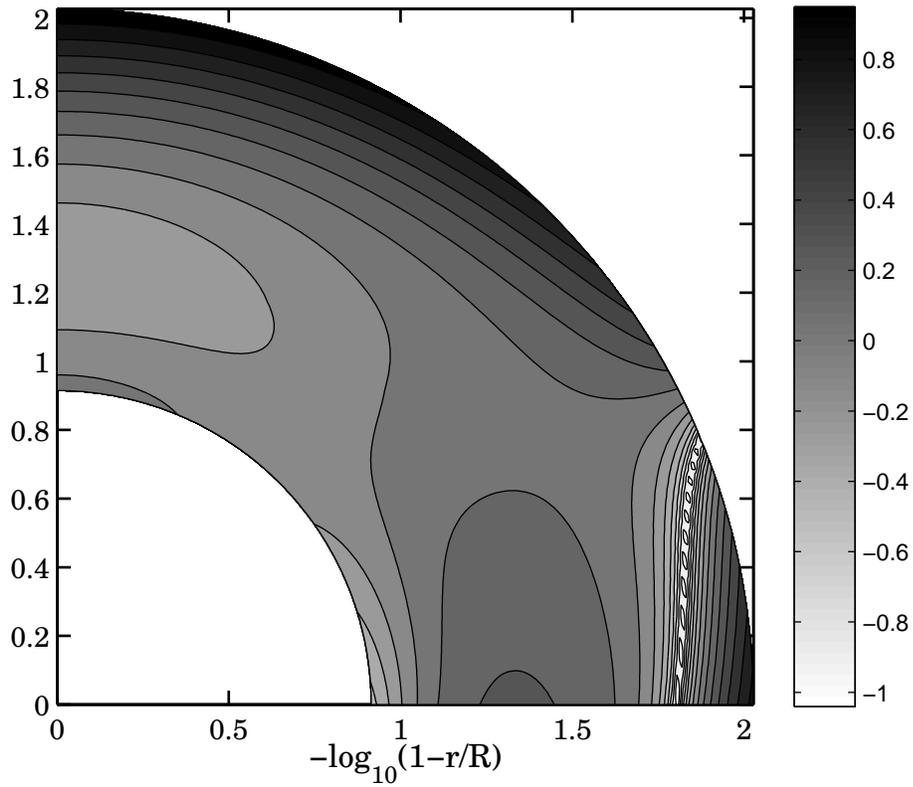}
\caption{\label{fig:sigma-2d} The distribution of (spindown portion of the) crustal strain
$\bar{\sigma}$ on a slice in a meridional plane, containing the
rotation axis (pointing upward) for model AU.  Results are normalized to 
$\Omega = c/R.$ Grayscale
indicates $\log_{10}\bar{\sigma}$. }
\end{figure}

\begin{figure}
\epsfig{file=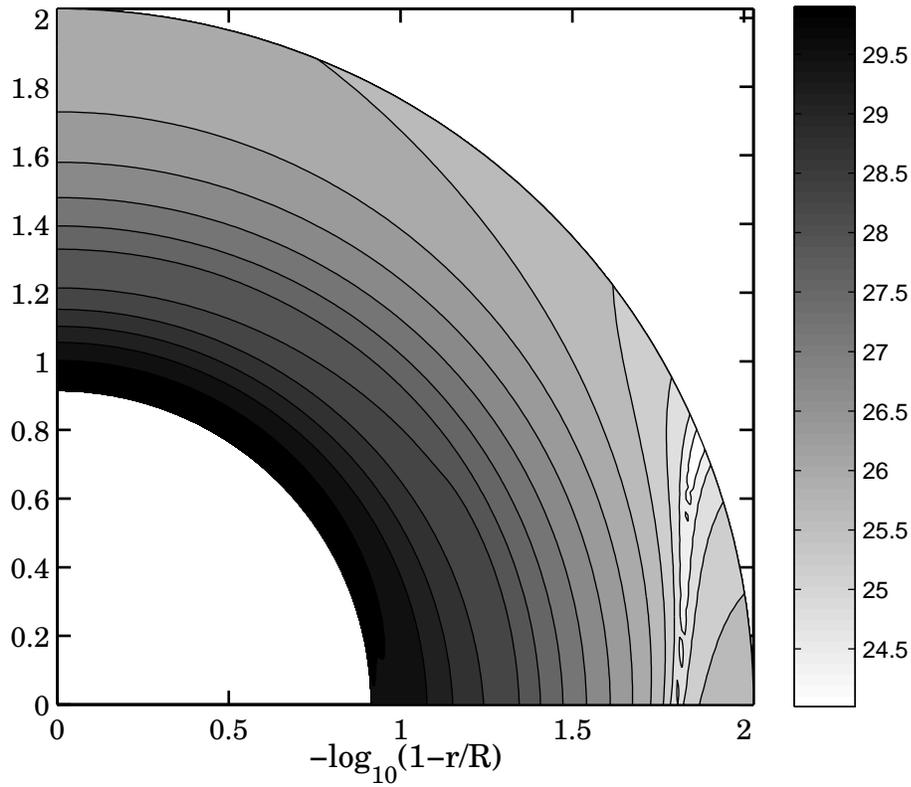}
\caption{\label{fig:tau-2d} The distribution of (the spindown portion of the) 
crustal shear stress
$\bar{\tau}$ on a slice in a meridional plane, containing the
rotation axis (pointing upward) for model AU. Results are normalized to 
$\Omega = c/R.$ 
Grayscale
indicates $\log_{10}\bar{\tau}$, where $\bar{\tau}$ is in cgs units
(erg~cm$^{-3}$). }
\end{figure}

\begin{figure}
\epsfig{file=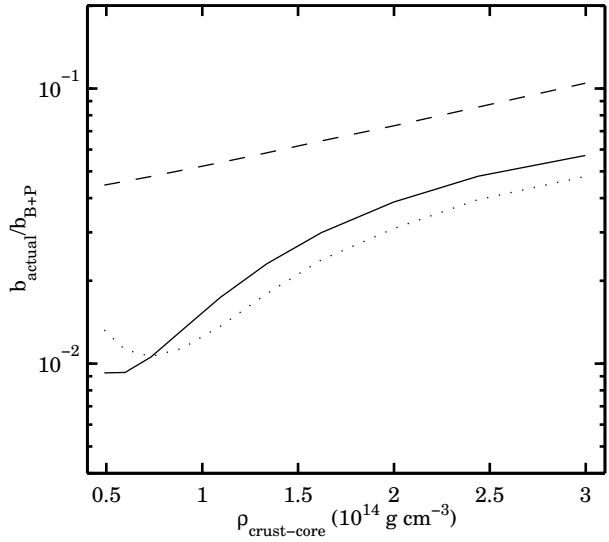}
\caption{\label{fig:b-vs-stdguess} The ratio of energy in crustal
stresses to that in centrifugal stresses, $b$, scaled to the
Baym-Pines estimate, plotted as a function of the location
of the crust-core boundary for equation of state AU (solid line), WS
(dotted line), and n1poly (dashed line).}
\end{figure}

\begin{figure}
\epsfig{file=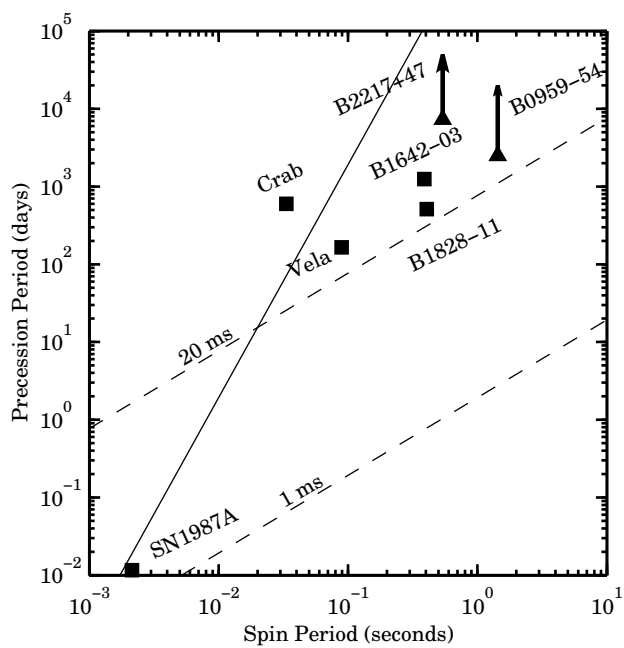}
\caption{\label{fig:prec-limit} Summary of precession candidates and
limits on their initial spins. Squares mark spin and precession
periods of candidates discussed in \S~6, while triangles with arrows
indicate lower limits on precession periods for B2217+47 and B0959-54.
The dashed lines indicate minimum precession periods of neutron stars
whose crusts last relaxed when they were spinning at 1~ms (lower line)
and 20~ms (upper line).  For a given initial spin period, all
precessing neutron stars must lie above the corresponding dashed
line. The solid line gives the precession period for NS's with relaxed
crusts. NS's that are slowing down should lie to the right of this line.}
\end{figure}

\begin{figure}
\plotone{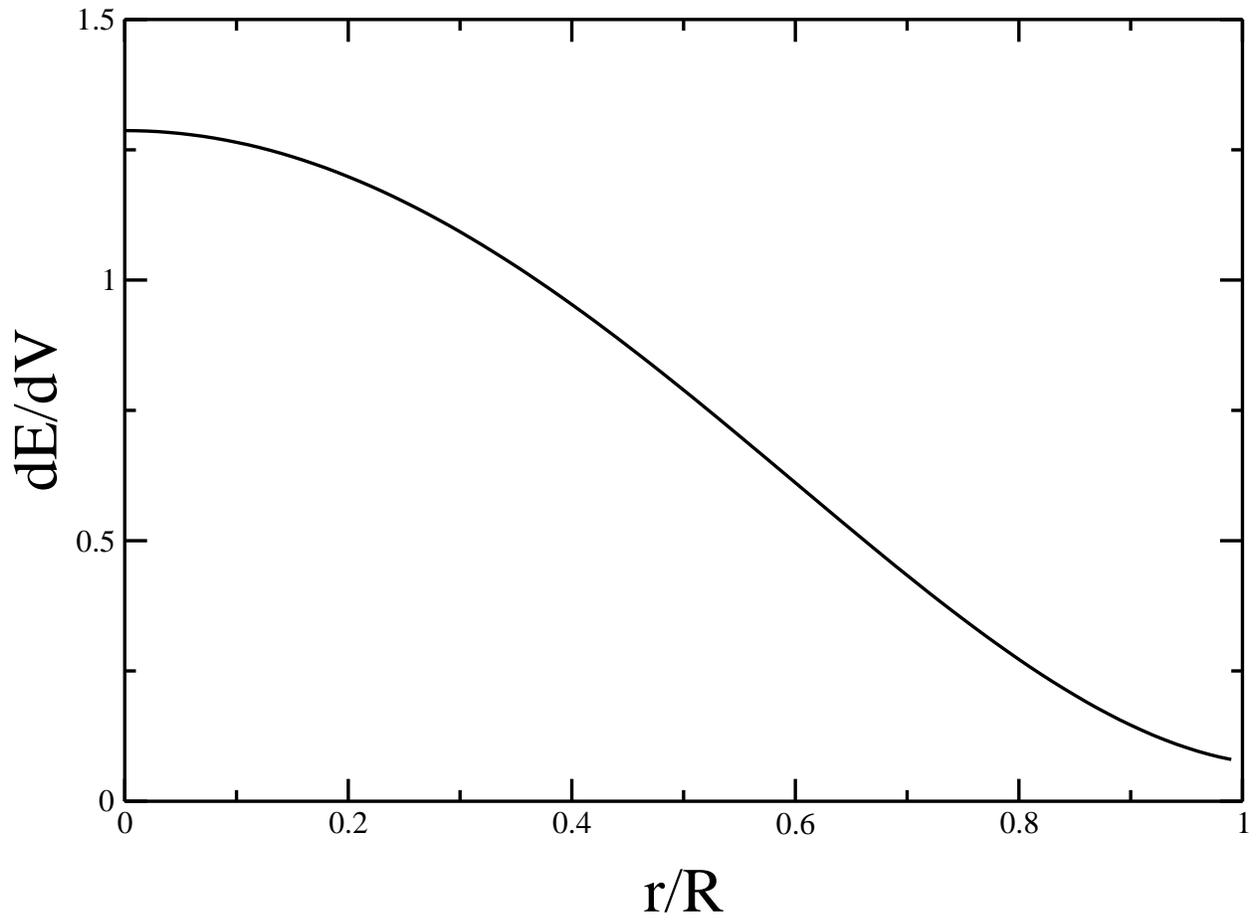}
\caption{\label{fig:strain_energy} The strain energy in
a homogenous, elastic sphere, as a function of r. Arbitrary units.}
\end{figure}

\begin{figure}
\epsfig{file=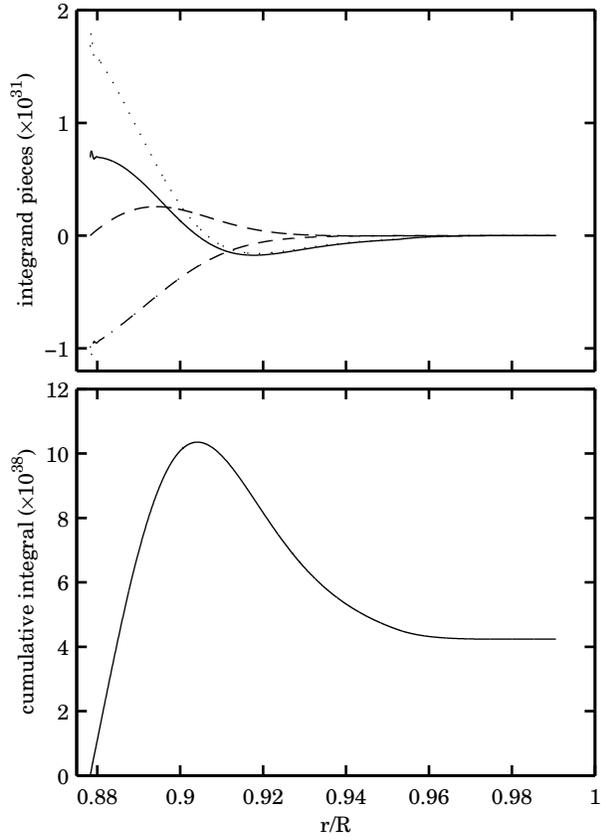}
\caption{\label{fig:pieces} Top panel: pieces of the integrand of
Eq.~(\ref{eq:qform}) proportional to $t_{rr}$ (dotted line),
$t_{r\perp}$ (dashed line), and $t_\Lambda$ (dot-dashed line),
computed for model AU. Bottom panel: cumulative integral.}
\end{figure}

\clearpage

\begin{deluxetable}{lllllllll}
\tabletypesize{\scriptsize}
\tablecaption{\label{table:eos-summary} 
Summary of NS models, computed $b$ values, and crustal strains}
\tablecolumns{9}
\tablewidth{5.7in}
\tablehead{
\colhead{EOS} &
\colhead{$R$} &
\colhead{$I_0$} &
\colhead{$I_C/I_0$} &
\colhead{$\Delta I_\Omega/I_0$} &
\colhead{$b$} &
\colhead{$b$} & 
\colhead{$\bar\sigma_{\rm pole}$} &
\colhead{$\bar\sigma_{\rm ave}$} \\
\colhead{} &
\colhead{(km)} &
\colhead{$\times10^{45}$g~cm$^2$} &
\colhead{$\times10^{-2}$} &
\colhead{$\times(\nu_s/$kHz$)^2$} &
\colhead{(actual)} &
\colhead{(B\&P)} 
}
\startdata
AU       & 12.13 &  1.11 & 0.86 & 0.259 & 2.47$\times10^{-7}$ & 9.20$\times10^{-6}$ &  17.1&  0.457 \\
UU       & 12.78 &  1.23 & 1.11 & 0.302 & 4.20$\times10^{-7}$ & 1.25$\times10^{-5}$ &  18.9&  0.495 \\
FPS      & 11.97 &  1.05 & 0.78 & 0.241 & 1.70$\times10^{-7}$ & 8.09$\times10^{-6}$ &  16.7&  0.432 \\
WS       & 12.05 &  1.10 & 0.70 & 0.257 & 1.47$\times10^{-7}$ & 7.62$\times10^{-6}$ &  14.9&  0.449\\
n1poly   & 12.53 &  1.15 & 2.86 & 0.275 & 1.31$\times10^{-5}$ & 2.14$\times10^{-4}$ &  103&  0.502\\
\enddata
\tablecomments{Summarizes our results on the crustal rigity and spindown strain
for 5 different EOS, and compares 
to the Baym-Pines estimate for $b$. 
The average value of the spindown strain
$\bar\sigma^{sd}_{ave}$ and its maximum value 
$\bar\sigma^{sd}_{pole}$ (attained on the spin axis, near the
North pole) are 
both normalized to $R\Omega/c = 1$. 
Results here assume the
density at the bottom is
$\rho_b = 1.5 \times 10^{14}{\rm  g/cm}^3$.}

\end{deluxetable}

\clearpage
\begin{deluxetable}{lllllllll}
\tabletypesize{\scriptsize}
\tablecaption{\label{table:candidates} 
Precession candidates: observations and inferences}
\tablecolumns{9}
\tablewidth{5.7in}
\tablehead{
\colhead{Name} &
\colhead{$P_s$} &
\colhead{$\dot{P_s}$} &
\colhead{$P_p$} &
\colhead{$\nu_{s,ref}$} &
\colhead{$\nu^{-1}_{s,ref}$} &
\multicolumn{2}{c}{Crustal Strain} \\
\colhead{} &
\colhead{(s)} &
\colhead{} &
\colhead{(days)} &
\colhead{(Hz)}&
\colhead{(ms)} &
\colhead{$\bar\sigma^{sd}_{pole}$} &
\colhead{$\bar\sigma^{sd}_{ave}$} &
}
\startdata

B1828-11&	0.405&		6$\times10^{-14}$ &		511 &	40&	26&	     
2$\times10^{-3}$&	5$\times10^{-5}$\\        
B1642-03&	0.388&		1.78$\times10^{-15}$&	1250		&	25&	41&	     
7$\times10^{-4}$&	2$\times10^{-5}$ \\     
B0833-45 (Vela)&	0.0893&		1.25$\times10^{-13}$&	165		&	32&	31&	     
10$^{-3}$&	3$\times10^{-5}$ \\    
B0531+21 (Crab)&	0.0334&		4.21$\times10^{-13}$&	600		&	10&	97&	$\sim$ 0     &	$\sim 0$ 	   \\    
B2217+47&	0.5384&		2.77$\times10^{-15}$&	$>$7300	&		$<$12&   $>$84&	     
$<$2$\times10^{-4}$&	$<$4$\times10^{-6}$ \\    
B0959-54&	1.437&		5.14$\times10^{-14}$&	$>$2500	&		$<$33&   $>$30&	     
$<$10$^{-4}$&	$<$3$\times10^{-5}$ \\    
SN1987A	&	2.14$\times10^{-3}$&    --	&		0.0116			&	597&	1.7& $\sim$0	     & $\sim$0 \\    

\enddata
\tablecomments{Summarizes our results for 6 precession 
candidates.
$\nu_{s,ref}$ is the crust's reference spin value, 
assuming crustal rigidity sets the observed 
precession period. 
The strain values are the current spindown strains
in the NS crust; i.e. the strains induced when a relaxed star 
spins down from the reference spin
$\nu_{s,ref}$ to the current spin $\nu_s$. For B2217+47 and 
B0959-54, only lower limits on the precession period are claimed in
the literature, yielding lower limits on $\nu_{s,ref}$. 
For the Crab and SN1987A, $\nu_{s,ref}$ is less than, or comparable to, 
$\nu_s$ for our fiducial
NS model, so there is no evidence of nonzero crustal spindown strain.
}

\end{deluxetable}


\begin{thebibliography}{}

\bibitem[]{}
Abney, M., Epstein, R. I. \& Olinto, A. V. 1996, ApJ, 466, L91.

\bibitem{ap85}
Alpar, A. \& Pines, D. 1985, Nature 314, 334 



\bibitem{ai75}
Anderson, P. W. \& Itoh, N. 1975, Nature, 314, 334. 

\bibitem{Baym_Pines_71}Baym, G. \& Pines, D. 1971, Ann. Phys., 66, 816


\bibitem{Cordes} Cordes, J. M. 1993  in{\it  Planets Around Pulsars}, 
ASP Conference Series, Vol. 36, eds: Phillips, Thorsett \& Kulkarni, p. 43 

\bibitem{cutler02}
Cutler, C. 2002; astro-ph/0206051

\bibitem{DAlessandro} 
D'Alessandro, F. \& McCulloch, P. M. 1997 \mnras, 292, 879

\bibitem{Deshpande} 
Deshpande, A. A. \& McCulloch, P. M. 1996, in {\it Pulsars:
Problems and Progress},  
ASP Conference Series, Vol. 105, eds.
S. Johnston, M.A. Walker, and M. Bailes., p. 101



\bibitem{Franco}
Franco, L.M.,  Link B., \& Epstein, R. I. 2000,  \apj 543, 987

\bibitem{jones00}
Jones, D. I. 2000, Ph.D. thesis, University of Wales, Cardiff

\bibitem{Jones_Andersson_2001}
Jones, D. I. \& Andersson, N. 2001, \mnras 324, 811

\bibitem{Jones_88}
Jones, P. B. 1988, \mnras 235, 545

\bibitem{Kaminiker_99}
Kaminker, A. D., Pethick, C. J., Potekhin, A. Y., Thorsson, V., \&
Yakovlev, D. G.  1999, \aa, 343, 1009

\bibitem{Link_Cutler}
Link, B. \&  Cutler C., 2002, MNRAS, 336, 211.

\bibitem{Link_Epstein_1996}
Link B. \& Epstein R.I. 1996, ApJ, 457, 844

\bibitem{Link_Epstein_2001}
Link, B. \& Epstein, R. I. 2001, \apj,  556, 392 

\bibitem[]{}
Link, B., Epstein, R. I., \& Lattimer, J. M. 1999, \prl, 83, 3362 

\bibitem{Lorenz_93}
Lorenz, C. P., Ravenhall, D. G., \& Pethick, C. J. 1993, Phys. Rev. Lett, 70, 379

\bibitem{Love}Love, A.E.H.  1944, The Mathematical Theory of Elasticity, 
Dover 4th edition, p.259

\bibitem{Lyne_88} 
Lyne, A. G., Pritchard, R. S. \& Smith, F. G. 1998, \mnras, 233, 667

\bibitem[]{}
Mendell, G. 1998, MNRAS, 296, 903

\bibitem[]{}
Mestel, L. \& Takhar, H. S. 1972, MNRAS, 156, 419

\bibitem[]{}
Mestel, L., Nittman, J., Wood, W. P., \& Wright, G. A. E. 1981, MNRAS,
195, 979

\bibitem{Middleditch}
Middleditch, J., Kristian, J. A., Kunkle, W. E., Hill, K.M., \& Watson, R. D. 2000; astro-ph/00010044

\bibitem{mm60}
Munk,  W. H., \& MacDonald, G. J. F.  1960, The Rotation of the Earth, 
Cambridge Univ. Press

\bibitem{Negele_73}
Negele, J. W., \& Vautherin, D. 1973, \nphysa, 207, 298

\bibitem[]{}
Nittmann, J. \& Wood, W. P. 1981, MNRAS, 196, 491

\bibitem{Oyamatsu_93}
Oyamatsu K.  1993, Nucl. Phys. A, 561, 431

\bibitem{ps72}
Pines, D., \& and Shaham, J. 1972,  Nature Physical Science, 235, 43; 
1972, Phys. Earth Planet. Interiors, 6, 103

\bibitem{Press_92}
Press,, W. H., Teukolsky, S. A., Vetterling, W. T., \& Flannery, B. P.
1992, NumericalRecipes in C, Cambridge Univ. Press 


\bibitem{RZC98} 
Ruderman M., Zhu T. Chen K. 1998, ApJ, 492, 267 

\bibitem{Shabanova}
Shabanova, T. V. 1990, Astron. Zh., 67, 536

\bibitem{Shaham_77}
Shaham, J. 1977, \apj, 214, 251

\bibitem{Shaham_86}
Shaham, J. 1986, \apj 310, 780 

\bibitem[]{}
Spitzer, L. 1958, IAU, Symp. 6: Electromagnetic Phenomena in Cosmical
Physics, 6, 169

\bibitem{Srohmayer91}
Strohmayer, T. E., van Horn, H.M., Ogata, S., 
Iyetomi, H., \& Ichimaru, S. 1991, ApJ, 375, 679 

\bibitem{sls00} Stairs, I. H., Lyne,  A. G. \& Shemar, S. L. 2000, 
Nature, 406, 484

\bibitem{Suleymanova}  
Suleymanova, S. A. \& Shitov, Y. P. 1994, \apj 422, L17

\bibitem{ucb00} 
Ushomirsky, G., Cutler, C., \& Bildsten, L. 2000, \mnras, 319, 902 

\bibitem[]{}
Wasserman, I. 2002, astro-ph/0208378

\bibitem{Wiringa_88}
Wiringa,  R. B., Fiks,  V., \& Fabrocini, A.  1988, Phys. Rev. C, 38, 1010

\end{thebibliography}
\end{document}